\documentclass{aa}
\usepackage{graphicx}
\usepackage{txfonts}
\usepackage{aalongtable}
\begin{document}
\newcommand{\kms}{$\mathrm{\,km\,s}^{-1}$}
\newcommand{\Vrot}{$\mathrm{V_{rot}}$}
\newcommand{\Vt}{$\mathrm{V_{t}}$}
\newcommand{\Teff}{${T_{\rm eff}}$}
\newcommand{\logg}{$\log \mathrm{g}$}
\newcommand{\FeH}{$\mathrm{[Fe/H]}$}
\title{Elemental abundances in the atmosphere of clump giants.
\thanks{Based on spectra collected with the ELODIE spectrograph at the
1.93-m telescope of the Observatoire de Haute Provence (France).}
\thanks{Full Tables \ref{tableA1} -- \ref{tableA4} are only available in
electronic form at http://www.edpsciences.org}
}
\author{T.V. Mishenina \inst{1},\, O. Bienaym\'{e} \inst{2},\, T.I. Gorbaneva
\inst{1},\, C. Charbonnel \inst{3,4}, C. Soubiran \inst{5},\,  \, S.A. Korotin
\inst{1},\, V.V. Kovtyukh \inst{1}}
\offprints{T.V. Mishenina, \\
           e-mail: tamar@deneb1.odessa.ua}
\institute{
Astronomical Observatory of Odessa
National University and Isaac Newton Institute of Chile,
Shevchenko Park, 65014, Odessa, Ukraine
\and
Observatoire Astronomique de l'Universit\'e Louis Pasteur, 11 rue de
l'universit\'e, F 67000 Strasbourg, France
\and
Geneva Observatory, CH 1290 Sauverny, Switzerland
\and
LATT CNRS UMR 5572, 14, av.E.Belin, 31400 Toulouse, France
\and
Observatoire Aquitain des Sciences de l'Univers, CNRS UMR 5804, BP 89,
33270 Floirac, France
}


\date{}
\authorrunning{Mishenina et al.}
\titlerunning{Clump giants}
\abstract
{}
{The aim of this paper is to provide the fundamental parameters and
 abundances for a large sample of local clump giants with a high accuracy.
This study is a part of a big project, where the
vertical distribution of the stars in the Galactic disc and
the chemical and dymamical evolution of the Galaxy are being investigated.}
{The selection of clump stars for the sample group was made applying a
color -absolute magnitude window to nearby Hipparcos stars.
The effective temperatures were estimated by the line depth ratio method.
The surface  gravities (\logg) were determined by two methods (the first one
was the method based on the ionization balance of iron and the second one was
the method based on  fitting of the wings of \ion{Ca}{i} 6162.17 \AA\ line).
The abundances of carbon and nitrogen  were obtained from molecular synthetic
spectrum, the Mg and Na  abundances were derived using the non-LTE
approximation. The ``classical" models of stellar evolution
without atomic diffusion and rotation-induced mixing were employed.
}
{The atmospheric parameters (\Teff, \logg, [Fe/H], \Vt) and Li, C, N, O, Na, Mg,
Si, Ca and Ni abundances in 177 clump giants of the Galactic disc were
determined. The underabundance  of carbon, overabundance of nitrogen
and ``normal" abundance of oxygen were detected.
A small sodium overabundance was found. A possibility of a selection of
the clump giants based on their chemical composition and the evolutionary
tracks was explored.
}
{The theoretical predictions based on the classical stellar evolution models are
in good agreement with the observed surface variations of the carbon and nitrogen
just after  the first dredge-up episode. The giants show the same behavior of
the dependencies of O, Mg, Ca, Si ($\alpha$-elements) and Ni (iron-peak element)
abundances vs. [Fe/H] as dwarfs do. This allows one to use such abundance ratios
to study the chemical and dynamical evolution of the Galaxy.
}

\maketitle
%

\section{Introduction}

Low-mass stars (i.e., below $\sim$ 2.3 $M_\odot$) climb the
red giant branch (hereafter RGB) with a degenerate He core whose mass
increases until it reaches a critical value of about  0.45 M$_{\odot}$
at the top of the RGB. At this point, helium ignites in a series of flashes
removing this degeneracy.
The star then becomes a ``clump" giant which undergoes the central He burning.
All the low-mass stars have similar core masses at the beginning of He burning,
and hence the similar luminosities.
Due to this fact, the red giants at this evolutionary stage exhibit a
specific feature in the color - magnitude diagram (CMD), called the ``clump".
According to evolutionary tracks and in agreement with the observation of
open clusters, all giants older than about 1 Gyr fall in the clump (Girardi
\cite{gir99}).

Clump giants are especially interesting from two aspects. Their intrinsic
brightness combined with the numerous and sharp features of their spectra
makes them good tracers for the galactic kinematics and chemistry. They are
also very useful in clarification of the advanced evolutionary stages of
the low mass stars. Until now, the comprehensive investigation
of the large sample of these stars has not been carried out yet.

This work is a part of our study of the Galactic disc surface mass
density (Siebert et al. \cite{siebet03},
Bienaym\'e et al. \cite{bienet06}) of the properties of both thin and
thick discs, and the abundance trends in the solar neighbourhood
(Soubiran et al. 2003, Mishenina et al. \cite{mskk04}, Soubiran \& Girard 2005).
In this part of the project, the local clump giants observed with high spectral resolution
and high signal-to-noise ratio (S/N) serve as reference stars. Here we take the
opportunity to provide  the strong observational constraints
for the theory of stellar evolution. As a matter of fact, the
chemical composition of the clump giant atmospheres reflects both the
chemical composition of the prestellar matter and nucleosynthesis
and mixing processes inside the stars. Therefore, for successful Galactic studies
we need to know the abundances of which chemical elements  are not affected by the
mixing processes. On the other hand, we can use the elements with abundances
affected by the mixing processes to distinguish different stages of the stellar
evolution, especially the clump phase.

While investigating the clump giants, we faced the problem of their selection
, since this region of the CMD is also occupied by the stars
of the ascending giant branch. The differentiation between the first-ascending
RGB stars and the ``clump" stars is rather complicated. Even for open cluster
stars, it is very difficult to establish with the good level of certainty,
which stars from the group under
investigation are the real clump ones (Pasquini et al., \cite{paset04}).
We can solve the problem of the correct selection using the extended
observational data on the clump giants, selected with  photometric criteria.
However, is it possible to identify the clump giants based on additional criteria,
including the data about their chemical composition?

When the star moves towards the RGB, the superficial convection zone deepens
and the nuclearly processed material penetrates into the atmosphere
changing of its chemical composition.
During this so-called first dredge-up phase, the surface abundances of Li, C,
N, and Na, together with the $^{12}$C/$^{13}$C ratio, are being altered.
The effect depends both on the stellar mass and metallicity (see
Charbonnel 1994).
Typically, the surface abundance of carbon decreases by $\sim$ 0.1-0.2 dex
and that of nitrogen increases by 0.3 dex or more (Iben \cite{iben91}).

Despite a large dispersion, the abundances of CNO elements and their isotopes
observed previously in the giants of the solar metallicity
(Lambert, Ries \cite{lamrie81}; Kj$\ae$rgaard et al. \cite{kijgus82})
were found to be in a good agreement with theoretical predictions (Iben \cite{iben91}).
We reinvestigate this problem with a larger sample of stars which have been
observed with higher quality.

For giants and supergiants of the solar metallicity the sodium overabundance
has been found in many studies (Cayrel de Strobel et al. \cite{cayet70},
Korotin \& Mishenina \cite{kormish99}, Boyarchuk et  al. \cite{boyet01},
Andrievsky et al. \cite{andet02}). As has been shown in some papers
(Mashonkina et al. \cite{maset93}, Korotin
\& Mishenina \cite{kormish99}) the Na overabundance cannot only be explained
by deviations from LTE. In this paper, we look into this problem using
recent theoretical considerations and track calculations.
The aim of this paper is to provide the fundamental parameters and
abundances for a large sample of local clump giants, determined with a high accuracy
and to use these data to 1) probe the vertical distribution of the stars in Galactic
disc, 2) investigate the Galactic chemical evolution, and 3) explore the possibility to
select the clump giants based on their elemental
abundances.

In Sect. 2 we described the photometric selection criteria for the clump giants
and the detail of our spectroscopic observations. In Sect. 3, the determination of
the atmospheric parameters is presented. In Sect. 4, the chemical abundances
 are determined. In Sect. 5, we discussed the behaviour of each element in our sample
with metallicity. In Sect. 6 we provided the analysis of the signs of
the first dredge-up in stars under current study.

\section{Selection of the stars and observations}

The  clump  stars  analysed in  this  paper  were  selected
using  a colour-absolute  magnitude window  applied to  nearby  Hipparcos
stars. Stars were  selected from the Hipparcos catalogue according to
the following criteria:

  $$\pi \ge 10\,\rm{mas}$$
  $$\delta_{\rm{ ICRS}} \ge -20\deg$$
  $$0.8 \le B-V \le 1.2$$
  $$ 0  \le M_{\rm V} \le 1.6 $$

where $\pi$ is the Hipparcos parallax and $
\delta_{\rm{ICRS}}$\footnote{The Hipparcos star positions are
expressed in the ICRS (see http://www.iers.org/iers/earth/icrs/icrs.html).}
the declination.
The Johnson $B-V$ colour was  transformed from the $Tycho$2
$B_{\rm  T}-V_{\rm T}$ colour applying Eq. 1.3.20 from \cite{esa97}:
  $$B-V  =  0.850  \,(B{\rm  _T}-V{\rm  _T})$$
The absolute magnitude $M_{\rm{V}}$  was computed using the $V$ magnitude
transformed from the Hipparcos $H_{\rm p}$ magnitude
to the Johnson system with the equation calibrated by Harmanec (\cite{har98}).

Among the selected nearly 400 giants, about a half of them were observed:
the first priority group contained stars, for which radial velocities were not
determined, and their [Fe/H] were not accurate or were old.
The known spectroscopic binaries were excluded.

The list was completed with a few clump stars having distances in the range
of 100-200\,pc, for which we were expecting the low metallicity.
Most of the clump stars with $B-V$ between 0.75 and 0.8, even lacking
previously published metallicities, were also included in the list and observed.
Finally, we  have included six  $Tycho$-2 stars that are  distant  clump
stars located  at about  500 pc from  the Galactic plane  toward the
North Galactic Pole.  They were previously identified from the low S/N
spectra (Bienaym\'e et al. \cite{bienet06}).

The spectra of the studied stars were obtained using the facilities of the 1.93\,m
telescope of the  Haute-Provence Observatoire (France)  equipped  with
\'echelle-spectrograph ELODIE. The resolving  power was 42000, the region
of the wavelengths was 4400 - 6800 \AA, the  signal-to-noise ratio
was about  130-230 (at  5500 \AA). The  initial processing  of the spectra
(image extraction, cosmic  particles removal,  flatfielding etc)
was  carried out  following  to Katz et  al.  (\cite{katzet98}).
Further processing  of the spectra (continuum level location, measurement
of  the  equivalent widths  etc)  was  performed  using the  software
package DECH20  (Galazutdinov \cite{gal92}). The  equivalent width were
measured using the Gaussian fitting.

\begin{figure}
\centering
\includegraphics[width=7cm]{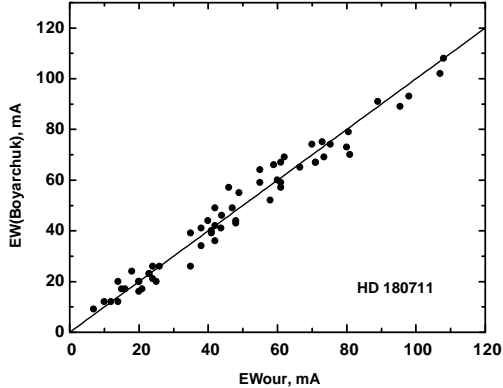}
\caption{ Comparison between EWs measured from the spectrum of HD 180711 and
those from Boyarchuk et al. (1996). Upper panel: EWs from Boyarchuk
et al.(1996); lower panel: our determinations. }
\label{f:EWcomp}
\end{figure}

In Fig. \ref{f:EWcomp} we have shown the comparison of our EWs measured in the
spectrum of HD 180711 with those reported by Boyarchuk et al. (\cite{boyet96}).
In the paper of Boyarchuk et al. (\cite{boyet96}) the spectra of program star
were obtained using the 2.6\,m  telescope of the Crimean astrophysical
observatory (Ukraine) with  coud\'e -- \'echelle spectrograph. The reciprocal
dispersion of those spectra  was 3 \AA/mm. An agreement between two
independent EW systems appears to be good, as one can see in
Fig.~\ref{f:EWcomp}. $\Delta$EW = EW(our) -- EW(Boyarchuk) = 0.23$\pm$4.5 m\AA.

The basic characteristics of studied stars are given in the Table~\ref{tableA1}.
The spectral classes $Sp$,  magnitudes $V$(Simbad) were taken from SIMBAD
database and magnitude $V$(Hipparcos) transformed fron the Hipparcos
$H_{p}$ to the Johnson system, parallaxes $\pi$ from Hipparcos catalogue
(ESA 1997), M$_{V}$ were calculated.
The 6 fainter stars of the sample are not part of the Hipparcos
catalogue. Their spectral type come from SIMBAD. The V magnitudes have
been extracted from the General Catalogue of Photometric Data by
Mermilliod et al (1997). Absolute
magnitudes Mv have been estimated from the TGMET software.

\section{Atmospheric parameters}
\subsection{Effective temperature \Teff}

The temperatures were determined with the very high level of accuracy
($\sigma = 10-15$ K)
using the line depth ratios The spectral lines of high and low excitation
 potentials respond differently to the change in effective temperature
($T_{\rm eff}$). Therefore, the ratio of their depths (or equivalent widths)
is a very sensitive temperature indicator.
This technique allows one to determine \Teff\, with an exceptional level of
accuracy. The method used is based on the ratio of the central
depths of two lines  having very different functional
 \Teff dependences. This method is independent of the
interstellar reddening and only marginally dependent on individual
characteristics
of the stars, such as rotation, microturbulence and metallicity.
NLTE effects will most likely affect ratios of high- and
low-excitation line strenthgs,
and ratios between different chemical elements.
Perhaps, these effects and also those
of varying individual stellar chemical abundance can be reduced by the
statistics of a large
number of different line ratios. The zero-point is well established and
is based on a
large number of independent measurements from the literature; it would
be unlikely that
the error on the zero-point is larger than 20--50 K.

\begin{figure}
\centering
\includegraphics[width=7cm]{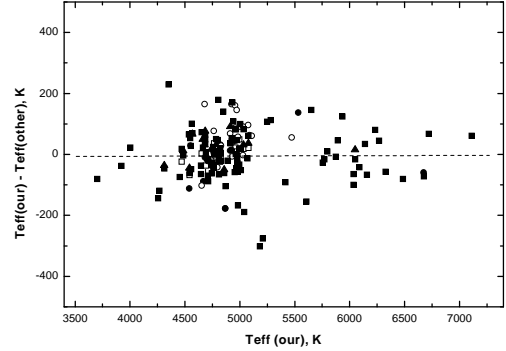}
\caption{Comparison between the temperatures derived in the present work and
 those derived by  Gray \& Brown (2001) -- {\it open squares},
  Blackwell \& Lynas-Gray (1998) -- {\it solid circles},
  Alonso, Arribas \& Mart\'{i}nez-Roger (1999) --  {\it solid triangles},
  Strassmeier \& Schordan (2000) -- {\it open circles},
  Soubiran, Katz \& Cayrel (1998) -- {\it solid squares}.
}
\label{f:Teff_comp}
\end{figure}

We used a set of 100 line ratio -- \Teff\, relations obtained
in Kovtyukh et al. (\cite{kovtet06}), with the mean random error
of a single calibration being 65--95 K
(45--50 K in most cases and 90--95 K in the least accurate
cases). The use of $\sim$70--100 calibrations per spectrum reduces
the uncertainty to 5--25 (1 $\sigma$) K.
This precision indicates that these 100 calibrations
are weekly sensitive to non-LTE effects, metallicity, surface gravity, micro-
and macroturbulence, rotation and other individual stellar parameters.
These relations have been calibrated with the reference
stars in common with  Gray \& Brown (\cite{gray01}) -- 21 stars, $\sigma$=27
K, Blackwell \& Lynas-Gray (\cite{BLG98}) -- 18 stars, $\sigma$= 81 K,
Alonso, Arribas \& Mart\'{i}nez-Roger (\cite{AAM99}) -- 14 stars,
$\sigma$=47 K, Strassmeier \& Schordan (\cite{ss00}) -- 20 stars,
$\sigma$=71 K and Soubiran, Katz \& Cayrel (\cite{soubet98}) -- 103 stars,
$\sigma$=106 K (see Fig.\ref{f:Teff_comp})
For the majority of stars, we obtained an internal error smaller than
20 K.

After the accurate effective temperatures have been determined, the other
atmospheric parameters were found iteratively.

\subsection{Surface gravity \logg, microturbulent velocity \Vt\ and
metallicity
[Fe/H]}

Because clump giants have similar luminosity but, different initial masses,
their gravities cannot be correctly determined using  their parallaxes and
mass data.
These \logg\ values are affected by $\pm$ 0.3dex if the stellar mass is
based on a assumption of mass of 2.2M$_{\odot}$.
We used two following methods of spectroscopic determinations
of the gravity \logg : 1) using the iron ionization equilibrium assumption,
where the average iron abundance determined from  \ion{Fe}{i}
lines and \ion{Fe}{ii} lines
must be identical, and 2) from  the wing fitting
of \ion{Ca}{i} 6162 \AA\, line.
For the method of the ionization equilibrium, we have used iron lines with EW
$<$ 120 m\AA; the wing profiles for such lines  practically do not depend on
damping constants, but
they are sensitive to microturbulent velocity \Vt$\,$ and metallicity [Fe/H].
Therefore, we take adopted \Teff\, and then we obtain the parameters (\logg,
\Vt\ and [Fe/H]) iteratively. Two or three steps were enough to get a
good convergence.

The second method is motivated by the fact that
the \ion{Ca}{i} line is strong in giants, and therefore its wing profile is sensitive to
the gas pressure in a stellar atmosphere, and therefore to the surface gravity.
The use of the \ion{Ca}{i} triplet lines (6102, 6122, 6162 \AA) as indicator of
surface gravity was proposed for dwarfs and subgiants by Edvardsson
(\cite{edv88}) and analyzed by Cayrel et al. (\cite{cayet96}). Cayrel et al.
(1996) have explored a possible influence of errors on the profiles of these
lines. They found that a change of 10$\%$ of the damping constants has a negligible
influence, a change of 15$\%$ becomes more or less detectable. The effective
temperatures were also varied by 100\,K and no alteration of
profiles was detected.
NLTE effects are important in the core of the line but negligible in the
wings. Recently Affer et al. (\cite{alfet05}) used this method  for K dwarfs and
subgiants. We have applied this method to giants to assess our gravity determination
in case of iron ionization  equilibrium. The \ion{Ca}{i} 6162 \AA \,
line which was recommended by Katz et al. (\cite{katzet03}) as a best luminosity
indicator among the triplet lines, was used.
We have estimated the influence of atmospheric parameter uncertainties on
accuracy of the gravity determination.
The \ion{Ca}{i} wings are not sensitive to the microturbulent velocity \Vt.
A change by 30 K in \Teff\, brings the errors in \logg\ about 0.05 - 0.10
(for \Teff = 5000 K and 4500 K, respectively).
To determine the calcium abundance value, we used weaker Ca I lines which are
presumed to be less affected by the damping and the microturbulent velocity.

\begin{figure}
\centering
\includegraphics[width=7cm]{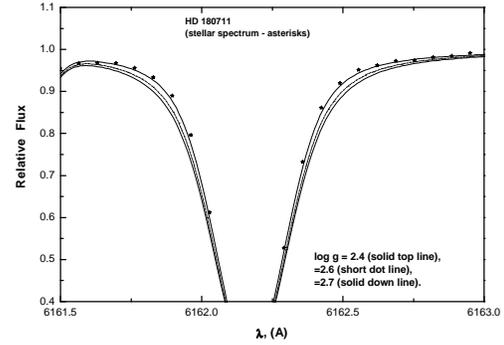}
\caption{Derivation of \logg\ from the \ion{Ca}{i} 6162 \AA\, profile for
HD 180711.}
\label{f:logg_ca}
\end{figure}

The departures from LTE in the computation of the wing profiles for these lines
 are negligible  for dwarfs and subgiants (Cayrel et al. \cite{cayet96}),
but in the case of giants it may elevate the level of uncertainty in \logg\ up
to 0.2 dex. The total error of the \logg\ determination  for giants
 is about 0.2-0.3 dex. An example of the line wing fitting for HD 180711
is given in Fig.~\ref{f:logg_ca}.  The values of \logg\, obtained by two methods,
are given in Table~\ref{tableA2}.
The mean difference $\Delta$\ (\logg(Ca)-\logg(IE))  is -0.01 $\sigma= \pm 0.09$
The results of these two methods applications are in good agreement.
The value of surface gravity for each star was obtained by keeping condition
of the ionization equilibrium between the Fe I and Fe II species and these values
were used for abundance determination.

The value of microturbulent velocity \Vt\ is determined by the standard
method from a condition of independence of the iron abundances determined
from the given line of \ion{Fe}{i} upon its equivalent width EW.
The accuracy of  \Vt\ determination is $\Delta$ \Vt =$\pm $ 0.2 km~s$^{-1}$.

The [Fe/H] metallicity is obtained from  the abundance determined from
\ion{Fe}{i} lines.
(In this paper we use the customary spectroscopic notation
[X/Y]=log$_{10}$(N$_X$/N$_Y$)$_{star}$ -- log$_{10}$(N$_{X}$/N$_{Y}$)$_{\odot}$)

In Table~\ref{tableA2} we give the mean \Teff, the number of calibrations used
($N$), the errors of the mean $\sigma$, \logg, \Vt, metallicities
 [Fe/H]$_{I}$ and  [Fe/H]$_{II}$ were determined from \ion{Fe}{i}
and \ion{Fe}{ii} lines, correspondingly.

\section{Determination of chemical abundances}

We employ the grid of stellar atmospheres from Kurucz (1993)
to compute abundances of Li, C, N, O, Na, Mg,Si, Ca and Ni.
The choice of the model was made using the standard
interpolation on \Teff\, and \logg.
The abundance analysis of  Si, Ca, Ni and Fe has been done
in the LTE approximation (Kurucz's WIDTH9 code) using the measured
equivalent widths of these elements'  lines and the solar oscillator strengths
(Kovtyukh \& Andrievsky \cite{kovand99}).
Abundances of Fe, Si, Ca, Ni were derived from a differential
analysis relatively to the Sun's data (see discussion in Mishenina et al. 2004).
In Table~\ref{tableA3} the relative-to-solar Fe, Si, Ca, Ni abundances and
individual errors are given.

\subsection{The Li abundance}
\begin{figure}
\centering
\includegraphics[width=7cm]{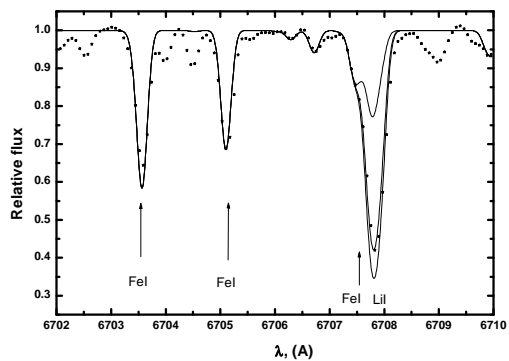}
\caption{Comparison between the observed specrum for HD 90633 and synthetic
one with the Li abundances logA(Li) = 1.0, 1.85, and 2.1.
 }
\label{f:li-synt}
\end{figure}

The Li abundances in program stars were obtained by fitting synthetic spectra
 to the observational profiles. We used STARSP LTE spectral synthesis code
developed by Tsymbal (\cite{tsy96}). Considering a wide range of temperatures
and metallicities of our sample stars, the special effort was put into
a compilation of a full list of atomic and molecular lines close to
the  $^{7}$Li 6707 \AA\ line (Mishenina \& Tsymbal, \cite{mistsy97}).
In Fig.~\ref{f:li-synt}, the comparison was made of the observed and the
 calculated spectra of HD 90633, for different lithium abundances log A(Li) = 1.0,
 1.85, and 2.1, where log A(X) = 12 + log(N$_{X}$/N$_{H}$).
 The derived values of log A(Li)$>$ 0.5 dex are given for 24 stars
in Table~\ref{tableA4}.
We consider the lithium abundance of about 0.5 dex  as a lower
limit of the reliable determination.
The comparison of our results with values found in the literature
(Brown et al. \cite{broet89}) shows a good agreement
$\Delta$ logA(Li)Brown - logA(Li)our = --0.01 $\pm 0.13$ (for 8 common stars).

\subsection{CNO abundances}

\begin{figure}
\centering
\includegraphics[width=7cm]{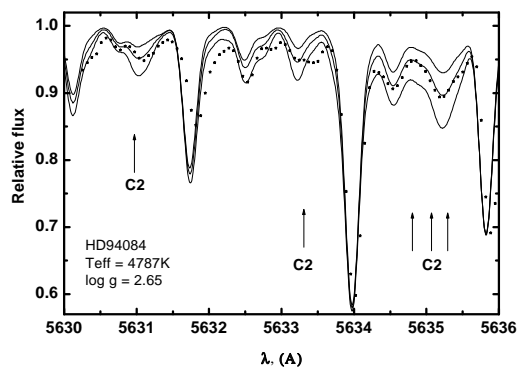}
\caption{Comparison between the observed specrum for HD 94084 and synthetic
one with the C abundances logA(C) = 8.36, 8.46, and 8.56.
 }
\label{f:c2}
\end{figure}

The abundances of carbon, nitrogen and oxygen are determined by the
method of  synthetic spectrum using the STARSP code (Tsymbal, \cite{tsy96}).
The spectrum of a molecule C$_{2}$ at the  5630 \AA\ (head of a band C$_{2}$
(0,1) of the Svan system  d$^{3}\Pi_{g}$ -- a$^{3}\Pi_{u}$ was used to derive
the carbon abundance.
The nitrogen abundance  was determined from the spectrum of a
molecule CN at 6330 \AA\, and 6470 \AA\, (red system
A$^{2}\Pi$ -- X$^{2}\Sigma$, heads of
bands CN (5,1) and (6,2)). The wavelengths  and parameters of molecular lines
(including  log gf) were taken from  Kurucz (1993), and they were
corrected using the technique proposed by Kuznetsova \& Shavrina
(\cite{kuzsha96}).
For these spectral regions,  the  contribution from blended lines of other
systems of a molecule C$_{2}$, molecule CN and  (NH, OH, CH, MgH and
SiH) was estimated. The lines from our list were compared to the lines
from solar spectrum, using the atmosphere model from the Kurucz's grid.  The
contribution from the  blended lines in the solar spectrum appears to be
insignificant (except for lines of a molecule CN in the regions of a
molecule C$_{2}$ and line [OI] 6300.3 \AA\,).  For the
region of a molecule C$_{2}$ and [OI] 6300.3 \AA\, line, the CN lines were
included in the final line list. The calculation was carried out with the following dissociation
potentials D$_{0}$  (C$_{2}$) = 6.15 eV and D$_{0}$ (CN) = 7.76 eV.
In Fig. \ref{f:c2}, a comparison between synthetic spectrum and observed one
near 5630 \AA\,  is  shown.
In  Table~\ref{tableA4} the abundances of carbon, nitrogen, oxygen are given
with the scale $log$ A(H) = 12. {\bf Below we use the values of relative to solar and iron abundances 
([C,N,O/Fe]) The solar C,N,O abundances are determined from fitting the synthetic and solar spectra. 
As solar spectra we used those of the Moon and asteroids that obtained with spectrograph ELODIE. 
The adopted solar values of log log A(C), log A(N), log A(O)  are  the following: 8.55, 7.97, 8.70, 
respectively.}

\subsection{NLTE abundances of magnesium and sodium}

In the spectra of cool giants the  lines of sodium and magnesium are
strong enough (EW$>$ 200 m\AA), therefore one can expect a significant
deviation  from LTE. For determination of the abundances of Na and Mg we
used NLTE approximation.  Four lines of \ion{Na}{i} and 9 lines of \ion{Mg}{i}
were considered.

NLTE abundances of Mg and Na were determined with the help of a modified
version of the MULTI code (Carlsson \cite{car86}) described in Korotin et
al. (\cite{koret99a}) and Korotin et al. (\cite{koret99b}). In such a modified
version, in particular, additional opacity sources from ATLAS9 code
(Kurucz \cite{kur93}) were included. This was done in order to
calculate the continuum opacity more precisely, and to
take into account the absorption by a great number of
spectral lines (especially within the region of the near UV).
It allows one to calculate more accurately the intensity
distribution in the region 900--1500 \AA. In turn,
this significantly affects the determination of the radiative
rates of bound -- free transitions. A simultaneous solution of
the radiative transfer and statistical equilibrium equations
has been performed in the approximation of complete
frequency redistribution for all the lines. All the NLTE
calculations were also based on the Kurucz's grid of
atmospheric models.

\subsection{Parameters of sodium and magnesium atoms}

The model of sodium atom as described by Sakhibullin
(\cite{sah87}), has been modified (see Korotin \& Mishenina
\cite{kormish99}). It consists of 27 levels of \ion{Na}{i} and the
ground level
of \ion{Na}{ii}. We considered the radiative transitions between
the first 20 levels of \ion{Na}{i} and the ground level of \ion{Na}{ii}.
Transitions between the remaining levels were used only
in the equations of particle number conservation. Finally,
46 $bound-bound$ and 20 $bound-free$ transitions were included in the
linearization
procedure. For 34 transitions the radiative rates were fixed.

\begin{figure}[hbtp]
\begin{center}
\includegraphics[width=7cm]{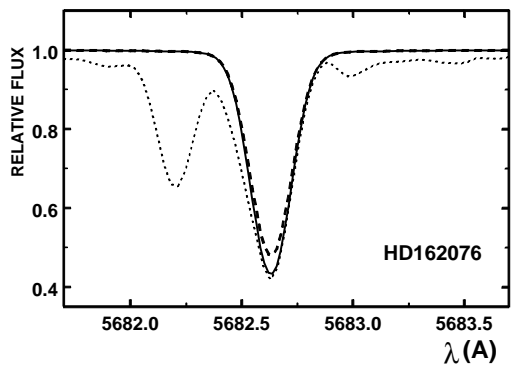}
\caption{NLTE profile fitting for HD 162076 (for the line \ion{Na}{i} 5683\AA)
and LTE profile (dashed line).
}
\label{f:na1-nlte}
\end{center}
\end{figure}

\begin{figure}[hbtp]
\begin{center}
\includegraphics[width=7cm]{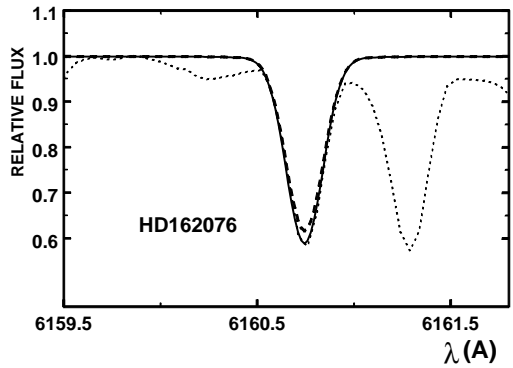}
\caption{Same as Fig.~\ref{f:na1-nlte} but for the lines \ion{Na}{i} 6164,
6160.}
\label{f:na2-nlte}
\end{center}
\end{figure}

We employed the model of magnesium atom consisting of 97 levels: 84 levels of
\ion{Mg}{i}, 12 levels of \ion{Mg}{ii} and a ground state of \ion{Mg}{iii}.
Within the described system of the magnesium atom levels, we considered the
radiative transitions between the first 59 levels of \ion{Mg}{i} and ground
level of \ion{Mg}{ii}. Transitions between the rest levels were not taken
into account and they were used only in the equations of particle number
conservation.
For detail see Mishenina et al., \cite{mskk04}.

\begin{figure}[hbtp]
\begin{center}
 \includegraphics[width=7cm]{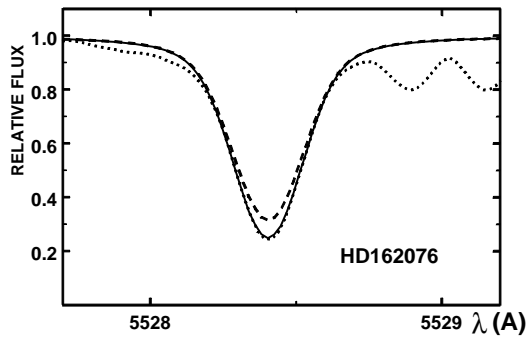}
\caption{Same as Fig.~\ref{f:na1-nlte} but for the lines  \ion{Mg}{i} 4703.
}
\label{f:mg1-nlte}
\end{center}
\end{figure}

\begin{figure}[hbtp]
\begin{center}
\includegraphics[width=7cm]{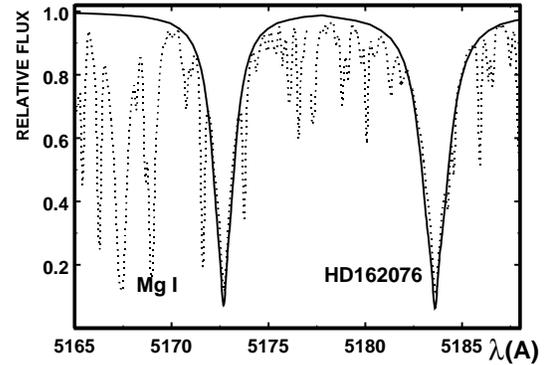}
\caption{Same as Fig.~\ref{f:na1-nlte} but for the lines  \ion{Mg}{i} 5173,
5184.}
\label{f:mg2-nlte}
\end{center}
\end{figure}

The difference between synthetic and observed spectra becomes visible if
the sodium and magnesium abundances are changed by about 0.05 dex.
The difference between sodium and magnesium abundances derived under the LTE assumption and
for NLTE case is within an interval of 0.10-0.15 dex. As an example, for
better comparison we have shown the LTE line profile (dashed line)in Figs. \ref{f:na1-nlte}, 
\ref{f:na2-nlte}, \ref{f:mg1-nlte}, \ref{f:mg2-nlte}.

In  Table~\ref{tableA4} NLTE abundances of Na and Mg are given in the scale
where logA (H) = 12.

\subsection{Abundance determination errors}

The metallicities [Fe/H] for  the giants have been determined. These determination
were based on the iron abundance value derived from \ion{Fe}{i} lines.
For this purpose, we used from 100 to 170 lines
depending on the temperature of the star: for the cooler stars the number of
iron lines was lower. The typical line-to-line scatter for \ion{Fe}{i} is 0.11
dex s.d.
The abundances of silicon, calcium and nickel have been determined from
12 to 22 lines of \ion{Si}{i}, 8 to 10 lines of \ion{Ca}{i} and 15 to 20 lines
of \ion{Ni}{i}. Typical standard deviations of the  abundances derived
from a single line of these elements are 0.12, 0.14, and 0.10 respectively.

Several factors may influence the abundance determination.
Among them  are: 1) the accuracy of the model parameters, 2) the equivalent
width measurements, 3) the quality of the synthetic spectrum adjustment,
and  4) internal errors of the method used. Concerning the last factor,
one can notice that somewhat different abundance results can be obtained if
one uses the LTE or Non-LTE approximations, 1D-, 2D- or 3D atmosphere models.
There are also uncertainties in atomic constants.
The use of the differential method minimizes these determination errors.
Uncertainties that are attributed to observed spectrum are the following.
A change in equivalent width of 2m\AA\ corresponds to a change in
abundance of about 0.03 -- 0.06 dex for \ion{Fe}{i},  \ion{Si}{i}, \ion{Ca}{i},
\ion{Ni}{i}. The fitting procedure between synthetic and observed spectra in
the case of C$_{2}$ lines produces uncertainty of about 0.02 dex. In other
cases (N, Na, Mg) it is about 0.05 dex. The value of total uncertainty due to
the choice of the stellar parameters is shown in Table \ref{tab1}. The
atmospheric parameters were changed by
$\Delta$~Teff = +100 K, $\Delta$~\logg = +0.2, $\Delta$~[Fe/H] = --0.25 for
[Fe/H] $<$ 0 and $\Delta$~[Fe/H] = +0.1 for [Fe/H]$>$0,
$\Delta$~Vt= +0.2 km s$^{-1}$.

As one can see from Table \ref{tab1}, the total uncertainty reaches 0.18 -- 0.24
dex for iron abundance determined from \ion{Fe}{ii} species and 0.10 -- 0.12 dex
in the case of the \ion{Fe}{i} species. For C, N, O abundances, such
uncertainties are: 0.13 -- 0.23, 0.09 -- 0.13, 0.09 -- 0.21
respectively.
In the case of carbon, the maximal error takes place for the cooler stars.
For oxygen the uncertainty is caused by the choice of the model metallicity
for metal-deficient stars. For Na, Mg, Si abundances, the total error is
about 0.08 -- 0.11 and for Ca and Ni it is  about 0.11 -- 0.16.
The microturbulence uncertaintly supplies the largest uncertaintly to the
\ion{Fe}{i} iron abundance.

\begin{table}
\caption[]{Abundance determination errors.
Parameter variation and corresponding uncertainty in
abundance determination ($\Delta$~Teff = +100 K, $\Delta$~\logg = +0.2,
$\Delta$~[Fe/H] = --0.25 for
[Fe/H] $<$ 0 and $\Delta$~[Fe/H] = +0.1 for [Fe/H]$>$0,
$\Delta$~Vt= +0.2 km s$^{-1}$)}.
\label{tab1}
\begin{tabular}{lccccc}
\hline
Element& $\Delta$\Teff & $\Delta$\logg & $\Delta$\FeH & $\Delta$\Vt &
$\Delta$
tot\\
\hline
\multicolumn{3}{c}{HD161178}&\multicolumn{3}{c}{4789/2.2/-0.24/1.3}\\
\hline
FeI   &   0.06&    0.01&   --0.03&   --0.10&    0.12 \\
FeII  & --0.10&    0.08&   --0.19&   --0.08&    0.24 \\
LiI   &   0.14&    0.00&     0.02&   --0.01&    0.14 \\
CI    & --0.12&    0.05&     0.03&     0.00&    0.13 \\
NI    &   0.08&    0.03&   --0.10&     0.00&    0.13 \\
OI    &   0.02&    0.09&   --0.19&     0.00&    0.21 \\
NaI   &   0.08&  --0.03&     0.02&   --0.07&    0.11 \\
MgI   &   0.07&  --0.02&   --0.04&   --0.06&    0.10 \\
SiI   & --0.02&    0.03&   --0.08&   --0.04&    0.10 \\
CaI   &   0.11&    0.00&     0.05&   --0.07&    0.14 \\
NiI   &   0.05&    0.03&   --0.06&   --0.08&    0.12 \\
\hline
\multicolumn{3}{c}{HD17361}&\multicolumn{3}{c}{4646/2.5/0.12/1.5}\\
\hline
FeI   &   0.03 &   0.03 &   0.03 &   --0.09&    0.10 \\
FeII  & --0.11 &   0.17 &   0.08 &   --0.05&    0.22 \\
LiI   &   0.16 & --0.01 &   0.00 &     0.00&    0.16 \\
CI    & --0.13 &   0.15 & --0.01 &     0.00&    0.20 \\
NI    &   0.03 &   0.07 &   0.05 &     0.00&    0.09 \\
OI    & --0.04 &   0.13 &   0.05 &     0.00&    0.14 \\
NaI   &   0.09 &   0.00 & --0.01 &   --0.07&    0.11 \\
MgI   &   0.05 &   0.01 &   0.03 &   --0.06&    0.08 \\
SiI   & --0.05 &   0.09 &   0.04 &   --0.05&    0.12 \\
CaI   &   0.09 & --0.01 & --0.01 &   --0.07&    0.11 \\
NiI   &   0.03 &   0.07 &   0.05 &   --0.11&    0.14 \\
\hline
\multicolumn{3}{c}{HD27697}&\multicolumn{3}{c}{4975/2.65/0.11/1.4}\\
\hline
FeI  &    0.07&    0.00&    0.00&    --0.08&    0.11 \\
FeII &  --0.07&    0.09&  --0.13&    --0.06&    0.18 \\
LiI  &    0.12&  --0.01&    0.07&    --0.01&    0.14 \\
CI   &  --0.10&    0.07&  --0.19&      0.00&    0.23 \\
NI   &    0.08&    0.03&    0.06&      0.00&    0.10 \\
OI   &    0.02&    0.09&    0.00&      0.00&    0.09 \\
NaI  &    0.07&  --0.02&    0.00&    --0.05&    0.09 \\
MgI  &    0.06&  --0.02&    0.03&    --0.03&    0.08 \\
SiI  &  --0.01&    0.03&  --0.05&    --0.04&    0.07 \\
CaI  &    0.09&  --0.01&    0.02&    --0.06&    0.11 \\
NiI  &    0.05&    0.02&  --0.05&    --0.14&    0.16 \\
\\
\hline
\hline
\end{tabular}
\end{table}

\section{Abundance trends with metallicity}

\subsection{The C, N, O  abundance}

\begin{figure}[hbtp]
\begin{center}
\includegraphics[width=7cm]{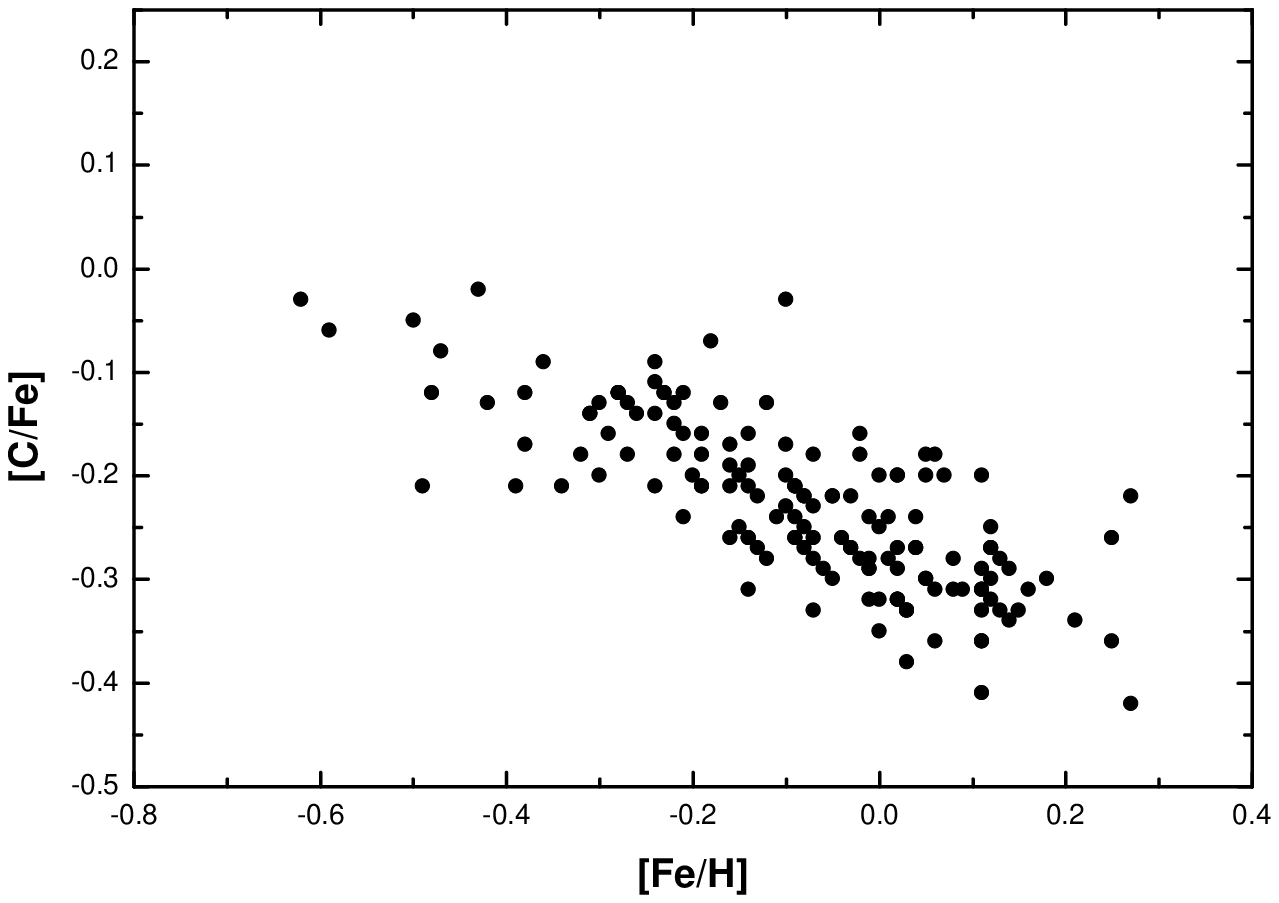}
\caption{Carbon abundance [C/Fe] vs. [Fe/H].
}
\label{f:c2-fe}
\end{center}
\end{figure}

\begin{figure}[hbtp]
\begin{center}
\includegraphics[width=7cm]{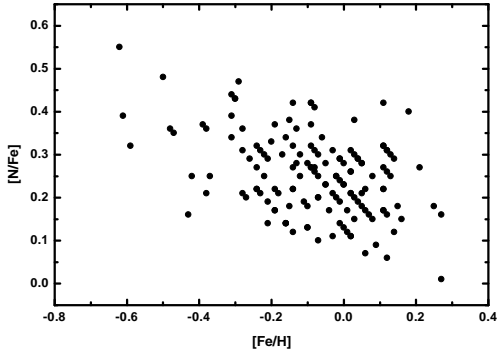}
\caption{Nitrogen abundance [N/Fe] vs. [Fe/H].
}
\label{f:N1}
\end{center}
\end{figure}

The abundance ratios [C/Fe], [N/Fe], [O/Fe] for each star in our set
are plotted against [Fe/H] in Figs.~\ref{f:c2-fe}, \ref{f:N1} and ~\ref{f:o-fe}.
For the whole sample of giants, the average values  of the abundances of
these elements are the following:
$<[C/Fe]>=-0.23\pm0.08$; $<[O/Fe]>= 0.08\pm0.16$; $<[N/Fe]>=0.25\pm0.09$,
for the stars of metallicity [Fe/H]$>$--0.3 they are:
$<[C/Fe]>=-0.24\pm0.07$; $<[O/Fe]>= 0.04\pm0.12$; $<[N/Fe]>=0.24\pm0.08$.
and for stars near solar metallicity --0.01 $<$[Fe/H]$<$ 0.01 they are:
$<[C/Fe]>=-0.28\pm0.05$; $<[O/Fe]>= 0.02\pm0.08$; $<[N/Fe]>=0.21\pm0.07$.
These averaged abundance ratios agree well with evolutionary model
predictions of Iben (\cite{iben91}), who showed that stars should have
decreased carbon and increased nitrogen abundances in their atmospheres.

\begin{figure}[hbtp]
\begin{center}
\includegraphics[width=7cm]{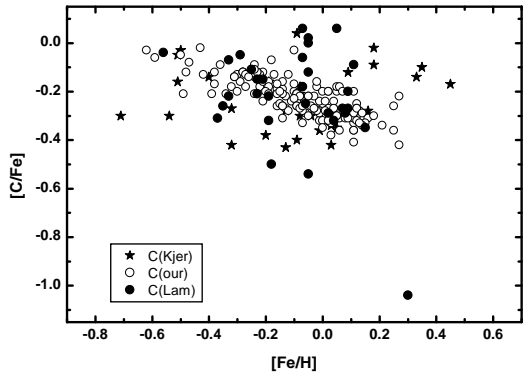}
\caption{Comparison of our  carbon abundances  (open circles)
 and  those of Kj$\ae$rgaard et al. 1982 (asterisks) and
 Lambert \& Ries 1981 (filled circles).
 }
\label{f:c-compar}
\end{center}
\end{figure}

\begin{figure}[hbtp]
\begin{center}
\includegraphics[width=7cm]{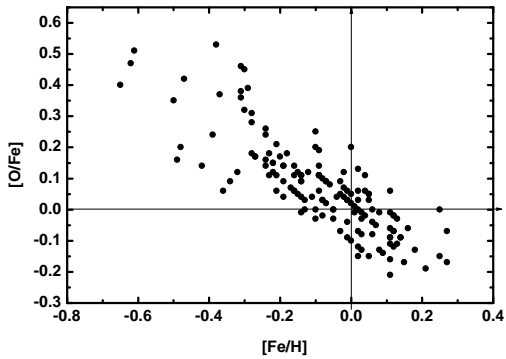}
\caption{Oxygen abundance [O/Fe] vs. [Fe/H].
}
\label{f:o-fe}
\end{center}
\end{figure}

Thus, our data (see  Fig.~\ref{f:c2-fe}) exhibit a clear
anticorrelation between [C/Fe] and [Fe/H].
In Fig.~\ref{f:c-compar} we compare our determinations (open circles)
of the carbon abundance with those
of  Lambert \& Ries \cite{lamrie81} (filled circles),
Kj$\ae$rgaard \& Gustafsson \cite{kijgus82} (asterisks) for all stars.
The mean values obtained in these mentioned works are:
$<[C/Fe]>=-0.22\pm0.21$
(Lambert \& Ries \cite{lamrie81}), $<[C/Fe]>=-0.31\pm0.30$
(Kj$\ae$rgaard \& Gustafsson \cite{kijgus82}). They are within the error
limits of determinations with our value $<[C/Fe]>=-0.23\pm0.08$, but
in our case we have  smaller scatter.
The dependence of [C/Fe] on [Fe/H] (see Fig. 12) is clearly observed only
for our clmp giant sample.
Whether is it a feature of the clump stars? Probably, not.
The same behaviour of [C/Fe] versus [Fe/H] was discovered in the disc dwarfs
(Bensby \& Feltzing \cite{benfel06}; Reddy et al. \cite{reddet06}), but
the avarege values of [C/Fe] are different for dwarfs and for giants.
Therefore, we can conclude that observed  trend is not a peculiar feature
of the clump giants, since it reflects the general tendency of the C abundance
decreasing with [Fe/H] increasing. We have detected it because our
[C/Fe] are obtained with a smaller scatter comparing to
others similar works.

We have found some (not distinctive) dependence between [N/Fe] and [Fe/H]
(see Fig.~\ref{f:N1}), but quite large scatter for our nitrogen data
prevents us from making a definitive conclusion. An analysis of the C and N abundances

within the frameworks of the evolutionary models is presented in Sec. 6.

The abundance of oxygen increases with the metallicity decrease
(see Fig.~\ref{f:o-fe}).
This behaviour is similar to $\alpha$-element behaviour in the dwarf stars.
This confirms the well-known fact that the relative-to-iron abundance of
the $\alpha$-elements decreases when the metallicity increases. This is connected
with the growing contribution from the SNe~I stars to the iron enrichment.
Obviously, the determined oxygen abundance in giants can be used
in an investigation of Galactic evolution.

\subsection{The Na abundance}

\begin{figure}[hbtp]
\begin{center}
\includegraphics[width=7cm]{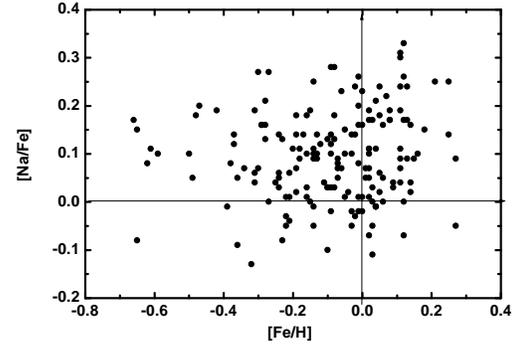}
\caption{Sodium abundance [Na/Fe] vs. [Fe/H].
}
\label{f:Na-fe}
\end{center}
\end{figure}

\begin{figure}[hbtp]
\begin{center}
\includegraphics[width=7cm]{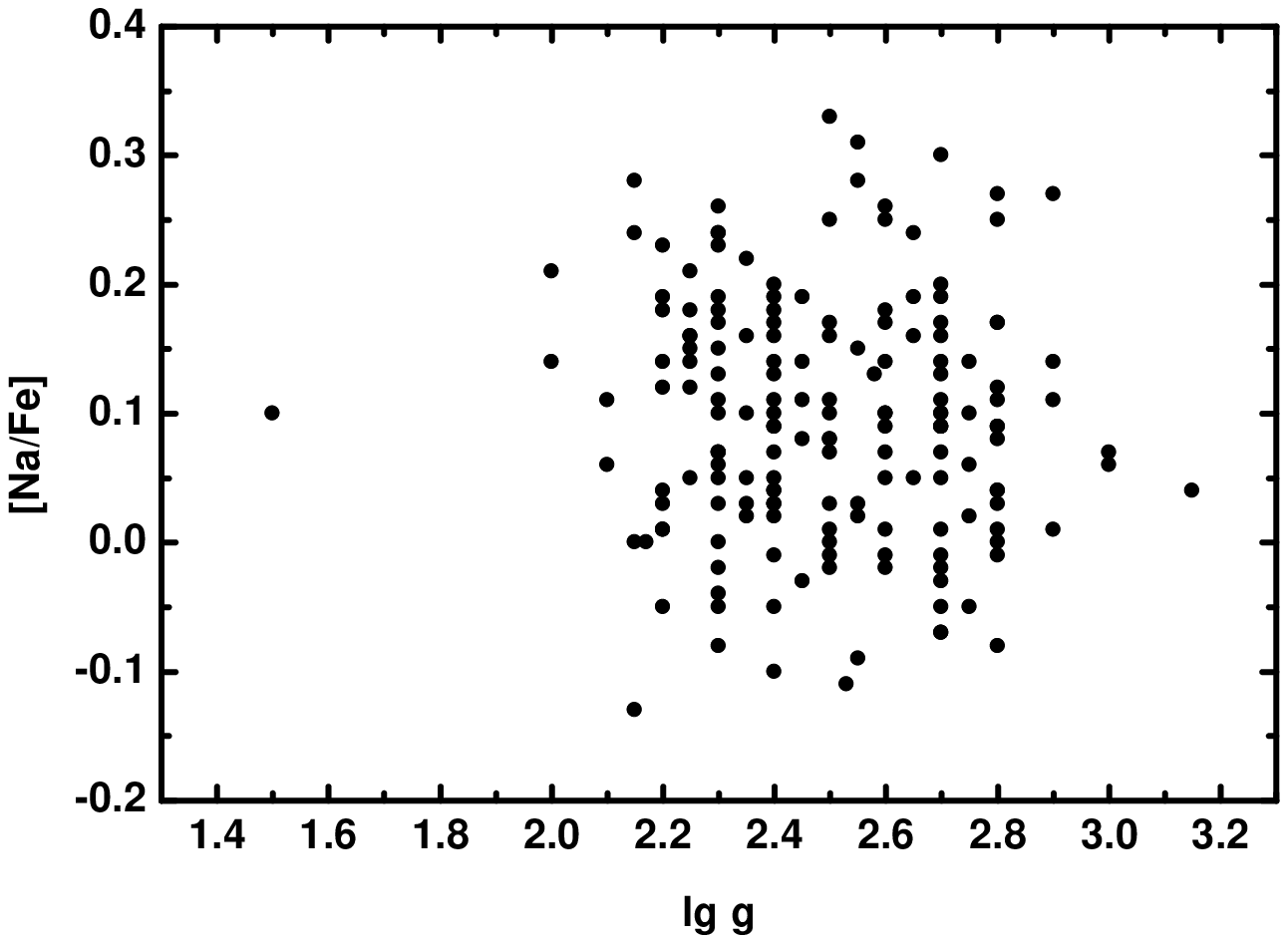}
\caption{Sodium abundance [Na/Fe] vs. \logg.
}
\label{f:Na-logg}
\end{center}
\end{figure}

\begin{figure}[]
\begin{center}
\includegraphics[width=7cm]{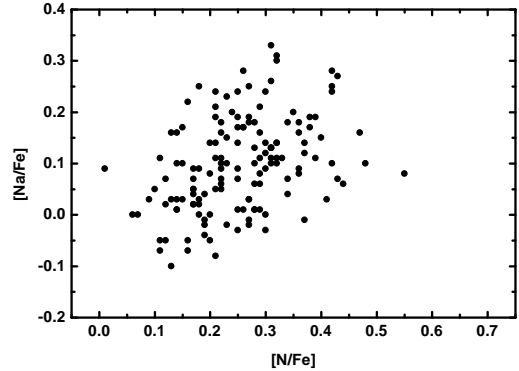}
\caption{Sodium abundance [Na/Fe] vs. nitrogen abundance [N/Fe]}
\label{Nafe-nfe}
\end{center}
\end{figure}

We found a small Na overabundance about 0.1 dex (see Fig.~\ref{f:Na-fe}),
and we also established that
there is no visible dependence of [Na/Fe] upon \logg \,
(see Fig.~\ref{f:Na-logg}). Some overabundance of sodium can be the sign
of the NeNa cycle opration. Nevertheless, an absence any
dependence between [Na/Fe] and \logg\, does not support this supposition.
From the other hand, this can be result of a restricted region of the
\logg\, we considered.
Additionally there is a correlation between [Na/Fe] and [N/Fe]
(see Fig.~\ref{Nafe-nfe}).
We notice that the behaviour of [Na/Fe] vs [Fe/H] is not
similar for giants and for dwarfs (Soubiran \& Girard \cite{sougir05}).
We will consider the sodium abundance below in Sec. 6.

\subsection{The $\alpha$-element and Ni abundances}

\begin{figure}[hbtp]
\begin{center}
\includegraphics[width=7cm]{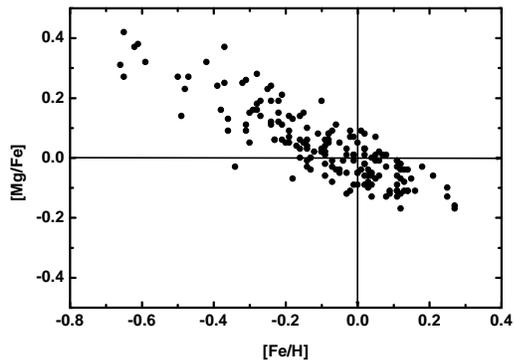}
\caption{Magnesium abundance [Mg/Fe] vs. [Fe/H]}
\label{f:mg}
\end{center}
\end{figure}

\begin{figure}[hbtp]
\begin{center}
\includegraphics[width=7cm]{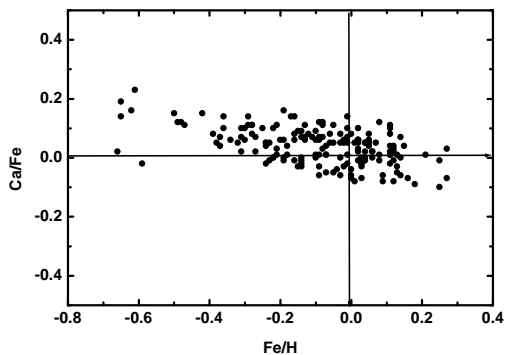}
\caption{Calcium abundance [Ca/Fe] vs. [Fe/H]}
\label{f:ca}
\end{center}
\end{figure}

\begin{figure}[hbtp]
\begin{center}
\includegraphics[width=7cm]{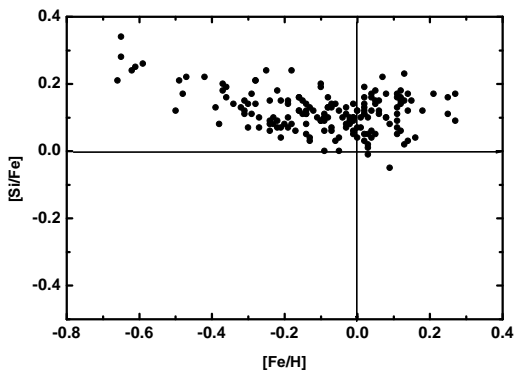}
\caption{Silicon abundance [Si/Fe] vs. [Fe/H] }
\label{f:si-fe}
\end{center}
\end{figure}

\begin{figure}[hbtp]
\begin{center}
\includegraphics[width=7cm]{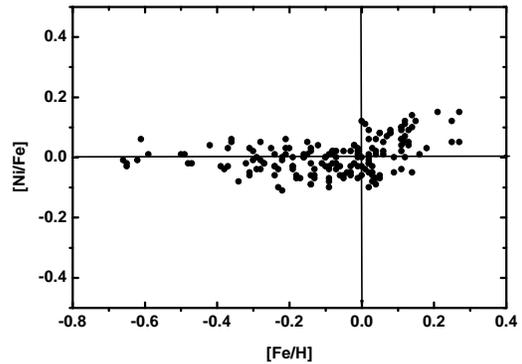}
\caption{Nickel abundance [Ni/Fe] vs. [Fe/H].}
\label{f:ni-fe}
\end{center}
\end{figure}

The behaviour of Mg, Ca, Si ($\alpha$-elements) and Ni (iron-peak element)
(see Figs.~\ref{f:mg}, \ref{f:ca}, \ref{f:si-fe}, \ref{f:ni-fe})
abundances vs. [Fe/H] in giants is the same as in dwarfs
(Soubiran \& Girard \cite{sougir05}).
It allows one to use abundances of these elements to study the chemical
and dynamical evolution of the Galaxy.

\section{Abundance variations due to the first dredge-up}

Due to surface abundance modifications during the first
dredge-up episode, clump giants do not exhibit the chemical pattern
that they inherited at their birth.
In this section we compare our data with first dredge-up theoretical
predictions.

\subsection{Stellar models}
We computed evolution models from the pre-main sequence up to the
AGB phase for stars with initial masses of 1.0, 1.5, 2.0, 2.5
and 3.0~M$_{\odot}$ and for three values of [Fe/H], i.e., --0.293, 0,
and +0.252 using the code STAREVOL (i.e., Siess
et al. 2000, Palacios et al. 2006). These are ``classical" models,
i.e., they do not take into account atomic diffusion and rotation-induced
mixing.
The nuclear reaction rates are those of NACRE (Angulo et al. 1999).
For the radiative opacities, we use the OPAL tables above 8000~K
(Iglesias \& Rogers 1996) and at lower temperatures the atomic and molecular
opacities of Alexander \& Fergusson (1994). The conductive opacities
are computed from a modified version of the Iben (1975) fit to
the Hubbard \& Lampe (1969) tables for non-relativistic
electrons and from Itoh et al.(1983) and Mitake et al.(1984)
for relativistic electrons.
The equation of state is described in detail in Siess et al.(2000) and
accounts for the non-ideal effects due to Coulomb interactions and pressure
ionization. {\bf The standard mixing length theory is used to model convection th $\alpha_{\rm MLT} = 1.75$ 
calibrated for the solar case. Neither overshooting, nor undershooting is considered for convection. The atmosphere 
is treated in the gray approximation and integrated up to an optical depth $\tau \simeq 5\times10^{-3}$.}
Mass loss is considered during the whole evolution and follow the Reimer's
(1975) empirical relation.

\subsection{Comparison with theoretical predictions}

\begin{figure*}[]
\begin{center}
\includegraphics[width=13cm]{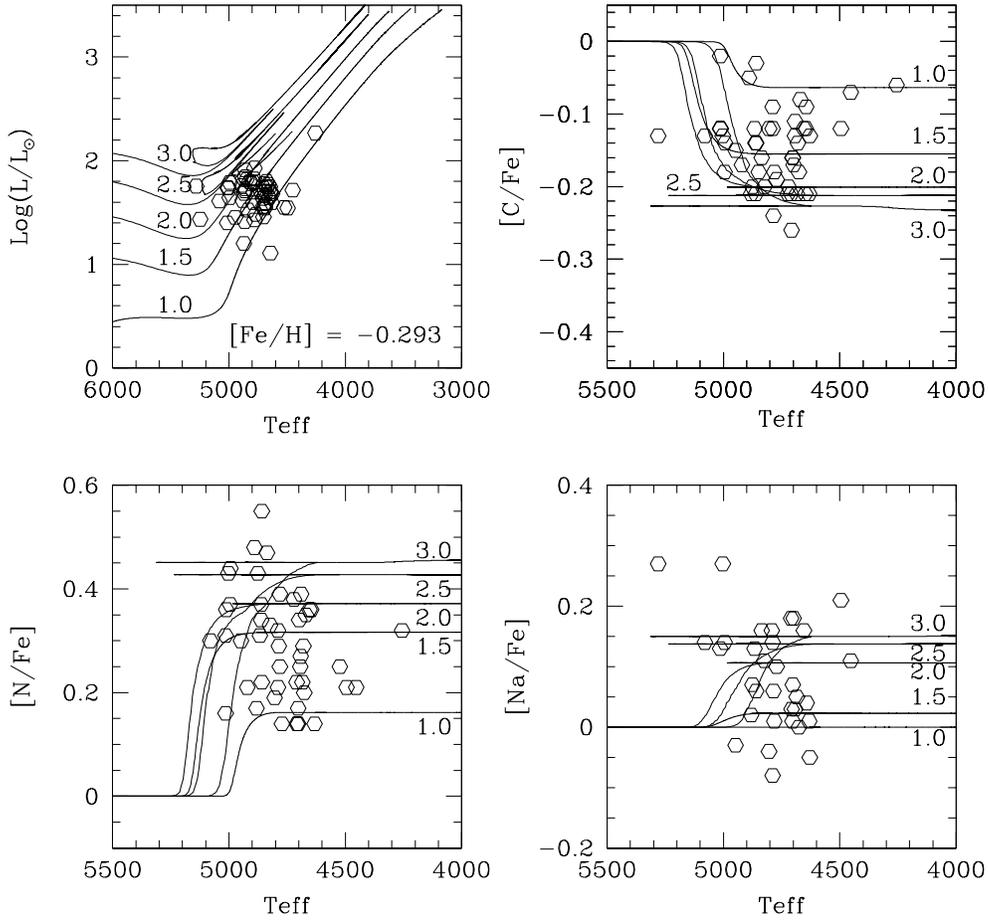}
\caption{Comparison of the theoretical tracks and predictions
for surface abundance variations with observations.
Position of our target stars with [Fe/H]$<$--0.15 in the H-R diagram}
\label{figureZ0092}
\end{center}
\end{figure*}

\begin{figure*}[]
\begin{center}
\includegraphics[width=13cm]{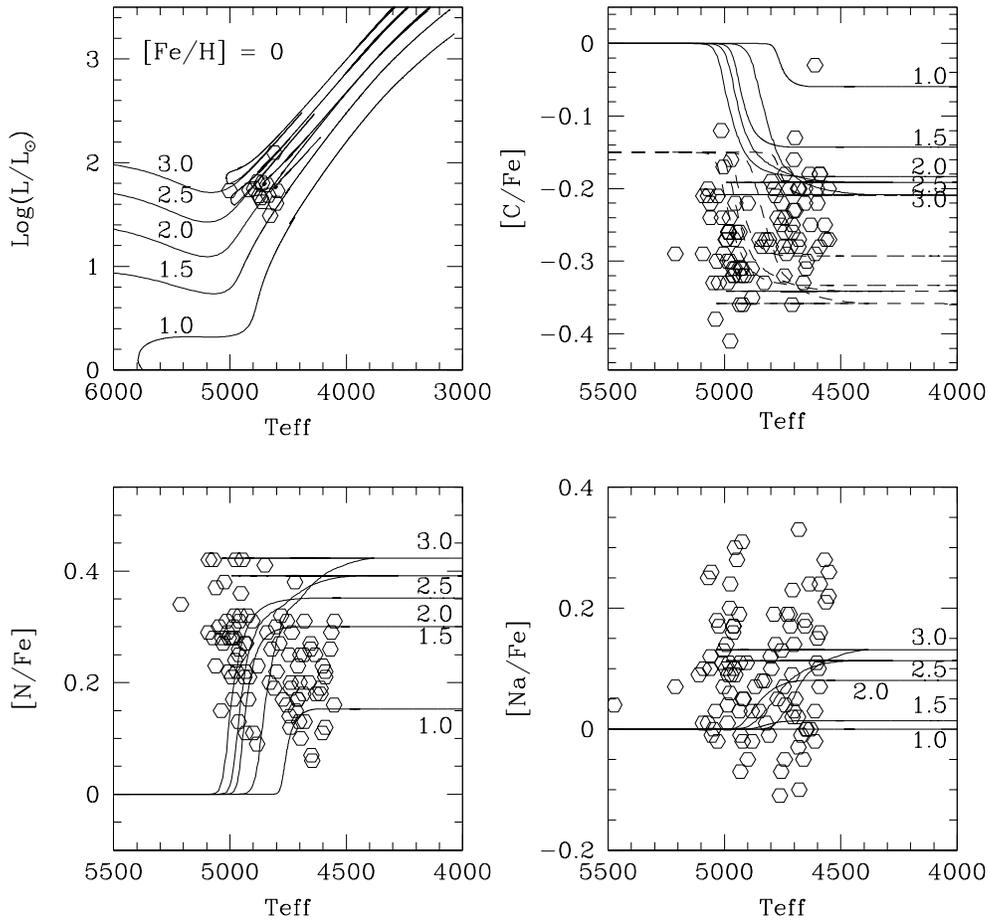}
\caption{Same as Fig.~\ref{figureZ0092} for [Fe/H]=0. Top right panel:
 the full lines are those assuming an initial [C/Fe]
 equal to solar, while the dotted lines are obtained by simply
 shifting the previous ones by --0.15.}
\label{figureZ02}
\end{center}
\end{figure*}

\begin{figure*}[]
\begin{center}
\includegraphics[width=13cm]{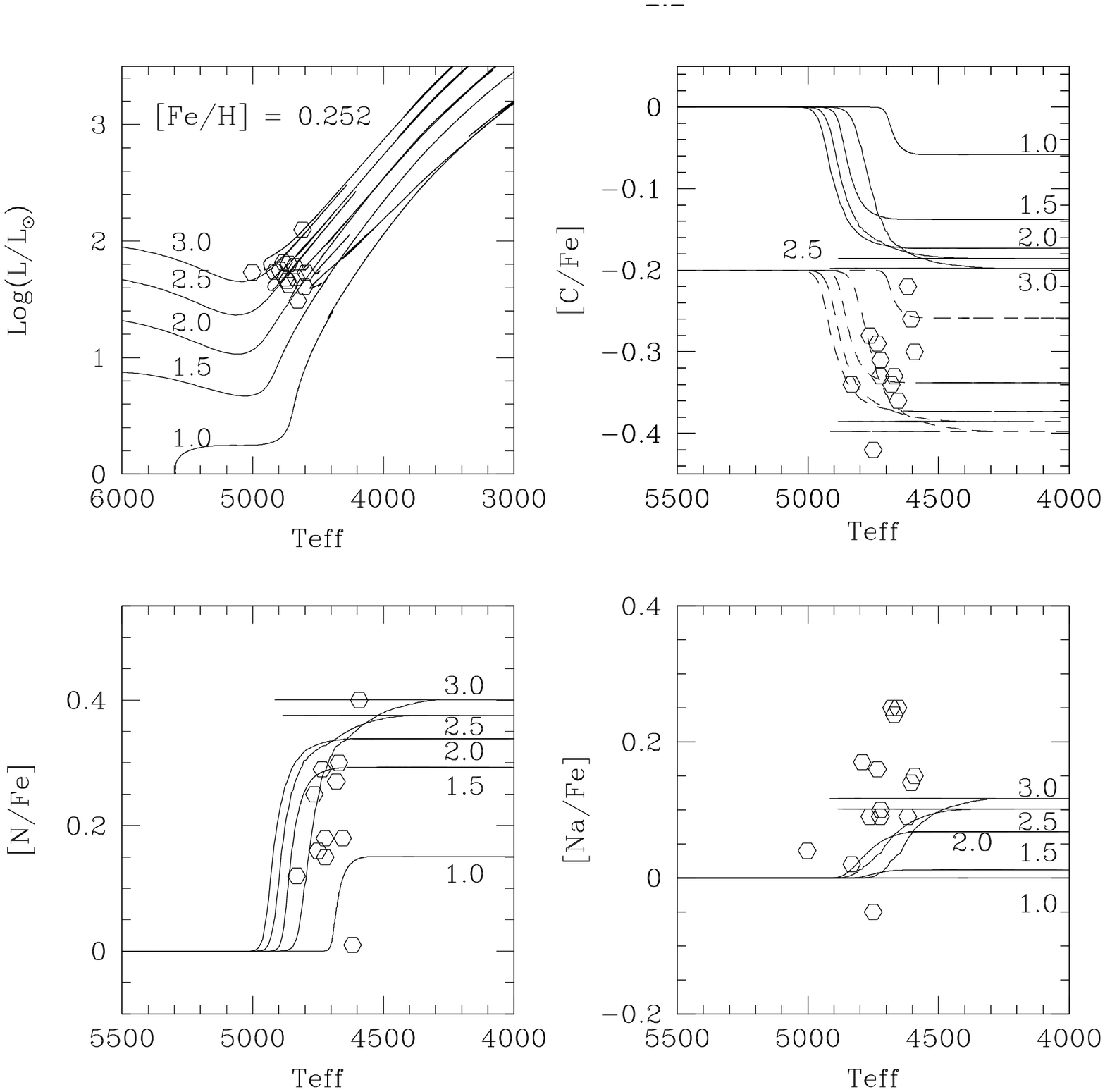}
\caption{Same as Fig.~\ref{figureZ02} for [Fe/H]=0.252 and [C/Fe]=--0.20.}
\label{figureZ03}
\end{center}
\end{figure*}

In Figs.~\ref{figureZ0092}, \ref{figureZ02} and \ref{figureZ03}
we show the corresponding evolutionary tracks together
with the positions of the sample stars in the HR diagram.
As expected, the objects appear to be slightly more massive
in the average when one moves to the higher metallicity.

Also shown in these figures are the predictions for the surface abundance
variations of C, N and Na as a function of effective temperatures
along the RGB together with the corresponding observational data.
Stars with [Fe/H] below --0.15 are compared with the [Fe/H]=--0.293 tracks,
those with [Fe/H] between --0.15 and +0.12 are compared with the [Fe/H]=0
tracks, and the more metallic ones with the [Fe/H]=+0.252 tracks.
As can be seen the region of the clump in the evolutionary tracks overlaps
the region where the first dredge-up ceases.

Our finding on the nitrogen abundance is in good agreement with the
prediction of the canonical theory of evolution for first dredge-up phase.

In the case of carbon, we show two sets of tracks for the two more metallic
subsamples. The full lines are those assuming an initial [C/Fe]
equal to solar, while the dotted lines are obtained by simply
shifting the previous ones by --0.15 and --0.20 dex respectively.
These quantities correspond to the values of the upper envelope of
[C/Fe] at the corresponding [Fe/H] (see Fig.~\ref{f:c2-fe}).
Again, the models explain well the observational pattern.

Regarding sodium the observed dispersion is higher than the theoretical one.
Numerous overabundances are observed, especially for the more metal-rich
subsample.
This cannot be attributed to an extra-mixing process, because
any additional processing of the envelope of the giant would also
lead to further changes in the C and N abundances which are not
observed in our sample.
One possibility to remove part of the discrepancy could lie in
the rates that intervene in the NeNa cycle. For the reaction
that forms sodium, $^{22}$Ne(p,$\gamma )^{23}$Na, the new rate calculated
by Hale et al. (2001) is slightly smaller (for the central temperature
of the models on the main sequence, i.e., below $\sim$ 20 million degrees)
than the NACRE prescription used in the present computations.
It would thus not favour stronger dredge-up of sodium.
On the other hand the present uncertainty of the
$^{23}$Na(p,$\gamma )^{24}$Mg and $^{23}$Na(p,$\alpha)^{20}$Ne reactions
is still large (Hale et al. 2004).
There also exist a possibility that the initial sodium abundance
was higher for some stars.
The disc dwarfs  show some dispersion of the
Na abundance and this is confirmed, for example, by Mishenina et al.(2003)
and Edvardsson et al. (1993) (especially for [Fe/H]$>$ 0).

\begin{figure}[]
\begin{center}
\includegraphics[width=7cm]{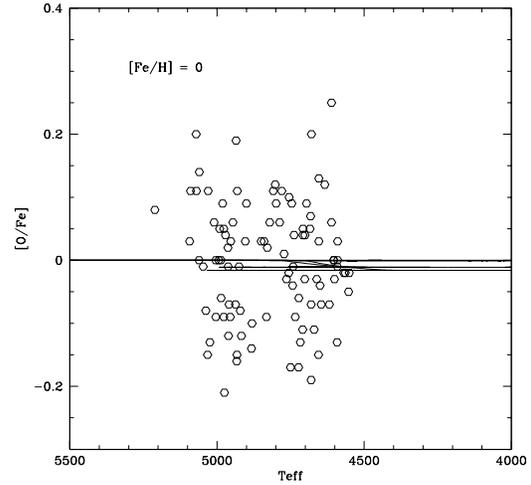}
\caption{Predictions for surface abundance variations of O during the
first dredge-up for the [Fe/H]=0 models and comparison with the
observations}
\label{OsurFevsTeffZ02}
\end{center}
\end{figure}

In  the  central regions  of  the  main  sequence stars,  $^{16}$O  is
partially  converted into  $^{14}$N. However  the  convective envelope
hardly reaches  the O-depleted region  during the first  dredge-up for
the mass  range considered here  (see for example Fig. 1 of Charbonnel
1994).
As a consequence  surface O variations are not  expected in our sample
stars (Fig.~\ref{OsurFevsTeffZ02}).
We just observe in our sample  of giants an O/Fe versus Fe/H variation
(Fig.~\ref{f:o-fe}) exactly similar to the variation observed with dwarfs (see
Figures 4 and 10 in Soubiran, Girard, 2004).

\subsection{The Li abundance}
\begin{figure}[hbtp]
\begin{center}
\includegraphics[width=7cm]{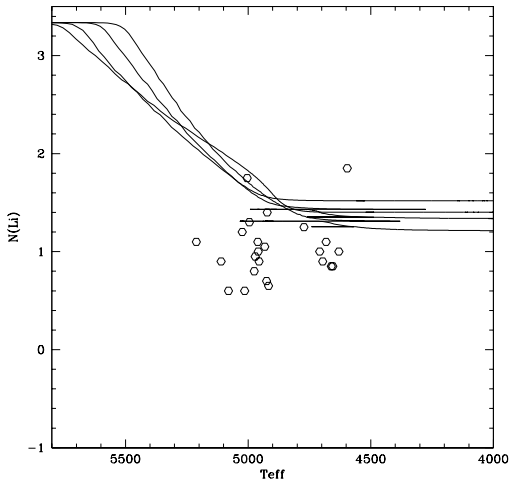}
\caption{Li abundance logA(Li) vs. \Teff\, for our sample stars
compared with the theoretical predictions for the tracks at [Fe/H]=0.
The line with the strongest Li depletion corresponds to the 1.0M$_{\odot}$
model}
\label{f:li-teff}
\end{center}
\end{figure}

According to the theory,  the surface Li abundance decreases with
respect to its value at the end of the main sequence (MS) by a factor from 30
to 60, depending on the stellar mass and metallicity (Iben 1991).
Starting from the present interstellar medium abundance of log N(Li) = 3.3
we thus expect after the first dredge up the Li values that lie around 1.5
as shown in Fig.~\ref{f:li-teff}.
Let's insist on the fact that these are ``classical" predictions which
do not take into account the effect of non-standard transport processes
such as those induced by rotation and which are thus not able to
explain the Li patterns observed in low-mass main sequence stars
(Charbonnel \& Talon 2005; see the review by Deliyannis et al. 2000)
By such, we have the right to expect for our giants lower values
of lithium, as proves to be true by observations
(Brown et al. \cite{broet89}, Mallik \cite{mal99}).

The observed Li abundances versus the effective temperatures for the giants
studied here are depicted in Fig.~\ref{f:li-teff}.

In the case of low-mass stars (M$_{star}<$2.2--2.5 M$_{\odot}$, HD 8733,
15453, 42341, 46374, 90633, 117304, 136138, 139254, 148604, 171994, 192836)
that undergo some extra-mixing at the RGB bump (Charbonnel et al. 1998,
Charbonnel \& Balachandran 2000), the fact that we
see some Li certainly indicates that these objects are RGB stars which
have not yet reached the bump. Otherwise their Li would have been
destroyed.

In the case of more massive stars which do not undergo this extra-mixing
because they do not go through the bump, Li should be consistent with
standard post dredge-up value predicted by the models.

Most likely, the Li abundance cannot be used as the criterion to segregate
the clump giants from RGB giants.

\subsection{Determination of the evolutionary status}

The stars considered in this study have been selected as clump stars
according to photometric criteria. Nevertheless, our sample could be
contaminated by ascending giant branch stars which cohabit with clump stars
in the considered region of the CMD
(see Figs.~\ref{figureZ0092}, \ref{figureZ02} and \ref{figureZ03}).
It would thus be interesting to perform more subtle separation of the
clump giants from the whole sample of the stars.

We aimed to do this selection by comparing the abundances of individual stars
with theoretical predictions of stellar evolution models.
This is however a very difficult task since the clump overlaps the region
where the first dredge-up ceases in the evolutionary tracks.
Despite this difficulty we checked the status of each sample stars
individually
following the procedure described below.

We first attributed a mass and evolutionary status to each object by
comparing its position in the HRD with the theoretical tracks.
The values of obtained masses for our target stars are given in the Table A1. 
Again, the stars with [Fe/H] below --0.15 are compared with the [Fe/H]=--0.293
models, those with [Fe/H] between --0.15 and +0.12 are compared with
the [Fe/H]=0 models, and the more metallic ones with the [Fe/H]=+0.252 ones.
From this first iteration 125 of our sample stars were identified
as possible RHB or clump stars, 4 as subgiants, 38 as probable RGBs,
and 2 are likely AGB stars.

Then we checked for each star whether its nitrogen abundance was
compatible with the model predictions for the corresponding stellar mass
attributed previously. The carbon abundance was used only as a cross-check
because of the variation it presents as a function of metallicity (see \S
5.1).
Among the 125 possible RGB/clump stars, 15 objects have no N determination
and 32 objects appear to still undergoing the first dredge-up dilution
as indicated by their [N/Fe] and effective temperature.
For the others we made the following distinctions :
(i) 38 stars are found to have completed their first dredge-up but present
N abundances slightly lower (by 0.05 to 0.2 dex)
than predicted values for the corresponding stellar mass. 3 stars have
N overabundances by $\sim$ 0.2 dex.
We consider however that these slight discrepancies are not significant
because of the observational errorbars on the effective temperature and
on the abundance determination.
Moreover, part of the small discrepancy can be accounted for by the fact that
not all the stars have the exact metallicity of the theoretical tracks
they are compared with.
(ii) 16 giants have a N abundance in good agreement with the post-dredge
predictions for the given stellar mass.
Both the (i) and (ii) stars would be preferentially
identified as RGB stars according to their effective temperature,
although they are still good clump candidates.
(iii) 21 stars could be selected as clump giants according to both their
N abundance in good agreement with the post-dredge predictions
and their effective temperature.

We have thus reliably selected 21 clump giants plus 54 clump candidates,
and about 100 usual giants that show all the signs of first dredge-up.
Unfortunately, we have to state that there exists some
uncertainty in the separation of the clump giants if we rely only on the
evolutionary tracks and elemental abundances that are sensitive to the
stellar evolution.

An important conclusion of the present study is that the theoretical
predictions of the classical models do account well for the
observed surface variations of both carbon and nitrogen during the
first dredge-up episode.

{\bf However, it  would be note that the considered analysis  and the result depend 
on accuracy of determination of chemical composition and  the used theoretical 
preconditions.}

\section{Conclusions}

We have performed the  detailed analysis of the atmospheric parameters
and the abundances of some elements in 177 giant stars.

The stars  analysed in  this study have  been selected as  clump stars
according to the photometric criteria.  We have estimated the possibillity
to  define  the  evolutionary status  of  these  giants  on the  basis  of
evolutionary tracks  and  their measured element abundances  that are
modified during  the stellar evolution. We reliably  selected 21 clump
giants,  about  54 clump candidates and about  100 usual giants that
show all the signs of first dredge-up.

The determined C, N  and Na abundances  in our program stars reflect
the CNO-  and NaNe cycle operation in the giant stars.

The O, Mg,  Ca,  Si   ($\alpha$-elements)  and  Ni  (iron-peak  element)
abundances  in  giants  show  trends  similar  to  those  observed  in
dwarfs. This allows one to use these abundances to study the chemical and
dynamical evolution of the Galaxy.

Acknowledgements. T.M. and V.K. want to thank the Observatoire Astronomique
 de l'Universit\'e Louis Pasteur de Strasbourg for kind hospitality. We also
thank the referee, Prof. B. Edvardsson, for very useful and fruitful comments
and suggestions on the manuscript. The work was made  within the framework
of the French - Ukrainian project "Dnipro" - "Egide".

\clearpage
\Online
\appendix

\section{}


\begin{longtable}{rlccrrr}
\caption[]{The basic characteristics of studied stars}\\
\hline
 Star& Sp & $V$ (Simbad)& $V$(Hipparcos)&$\pi$ (0.001)&$M_{V}$ &Masses\\
\hline
\endfirsthead
\caption[]{Continued.}\\
\hline
 Star& Sp & $V$ (Simbad)& $V$(Hipparcos)&$\pi$ (0.001)&$M_{V}$ & Masses\\
\hline
\endhead
\hline
\endfoot
\label{tableA1}
  2910& K0III   &5.347 & 5.38 &  12.69 &  0.432 &2.2   \\
  4188& K0III   &4.775 & 4.77 &  15.54 &  0.331 &2.5   \\
  4482&  G8II   &5.515 & 5.51 &  12.45 &  0.646 &2.5   \\
  5395&G8III-IV &4.632 & 4.62 &  15.48 &  0.202 &2.0   \\
  6319& K2III:  &6.193 & 6.20 &  10.01 &  0.696 &1.7   \\
  6482& K0III   &6.101 & 6.09 &  10.58 &  0.780 &2.0   \\
  7106&K0,5III  &4.523 & 4.51 &  20.11 &  0.563 &2.0   \\
  7578& K1III   &6.050 & 6.04 &  10.36 &  0.647 &2.0   \\
  8207& K0III   &4.875 & 4.87 &  16.68 &  0.550 &2.5   \\
  8599& G8III   &6.167 & 6.17 &  12.17 &  1.176 &1.2   \\
  8733&   K0    &6.450 & 6.44 &  10.71 &  1.262 &2.0   \\
  9408& G9III   &4.692 & 4.68 &  15.96 &  0.303 &2.0   \\
 10975& K0III   &5.947 & 5.94 &  10.57 &  0.705 &1.5   \\
 11559& K0III   &4.621 & 4.61 &  17.11 &  0.471 &2.7   \\
 11749& K0III   &5.698 & 5.69 &  10.18 &  0.257 &2.0   \\
 11949&  K0IV   &5.712 & 5.70 &  13.25 &  0.861 &1.0   \\
 15453& K2III   &6.098 & 6.09 &  11.4  &  0.913 &1.7   \\
 15755& K0III   &5.846 & 5.84 &  12.54 &  0.813 &1.5   \\
 15779& G3III   &5.364 & 5.36 &  12.27 &  0.414 &2.5   \\
 16247& K0III:  &5.819 & 5.81 &  11.18 &  0.549 &1.0   \\
 16400& G5III:  &5.656 & 5.65 &  10.29 &  0.334 &2.5   \\
 17361&K1,5III  &4.510 & 4.52 &  18.06 &  0.292 &2.0   \\
 18885& G6III:  &5.834 & 5.84 &  11.55 &  0.693 &2.2   \\
 19270& K3III   &5.648 & 5.64 &  10.21 &  0.240 &2.5   \\
 19787& K2III   &4.350 & 4.35 &  19.44 &  0.405 &2.7   \\
 19845& G9III   &5.921 & 5.93 &  10.47 &  0.684 &2.5   \\
 20791&G8.5III  &5.690 & 5.70 &  11.23 &  0.630 &2.5   \\
 25602&K0III-IV &6.320 & 6.31 &  10.21 &  0.894 &1.2   \\
 25604& K0III   &4.354 & 4.36 &  18.04 &  0.207 &2.7   \\
 26546& K0III   &6.091 & 6.09 &  11.42 &  0.939 &2.0   \\
 26659& G8III   &5.477 & 5.47 &  10.74 &  0.397 &3.0   \\
 26755& K1III   &5.727 & 5.72 &  12.38 &  0.679 &1.5   \\
 27348& G8III   &4.933 & 4.93 &  14.42 &  0.423 &3.0   \\
 27371& G8III   &3.654 & 3.65 &  21.17 &--0.043 &3.0   \\
 27697& G8III   &3.746 & 3.77 &  21.29 &  0.070 &3.0   \\
 28292& K2III   &4.971 & 4.96 &  16.78 &  0.458 &1.0   \\
 28305& K0III   &3.540 & 3.53 &  21.04 &--0.185 &3.0   \\
 28307& G7III   &3.847 & 3.84 &  20.66 &  0.100 &3.0   \\
 30557& G9III   &5.642 & 5.64 &  10.16 &  0.287 &2.5   \\
 31444&G6/G8III &5.726 & 5.71 &  11.3  &  0.719 &2.2   \\
 33419& K0III  & 6.118 & 6.11 &  10.37&   0.735 &2.0   \\
 33618&K2III-IV& 6.147 & 6.15 &  10.61&   0.738 &1.5   \\
 34200&   G5   & 6.384 & 6.39 &   8.5 &   0.749 &2.5   \\
 34559& G8III  & 4.956 & 4.96 &  15.83&   0.652 &2.5   \\
 35369& G8III  & 4.144 & 4.13 &  18.71&   0.167 &3.0   \\
 37638& G5III: & 6.168 & 6.17 &  10.26&   0.957 &2.5   \\
 39070& G8III  & 5.490 & 5.49 &  10.51&   0.364 &3.0   \\
 39910& K2III: & 4.874 & 5.87 &  10.71& --0.497 &3.0   \\
 40020& K2III  & 5.891 & 5.89 &  10.62&   0.536 &2.2   \\
 40801& K0III  & 6.092 & 6.08 &  12.47&   1.106 &1.0   \\
 42341& K2III  & 5.563 & 5.56 &  15.53&   1.023 &2.0   \\
 43023& G8III  & 5.835 & 5.83 &  10.36&   0.603 &2.5   \\
 45415& G9III  & 5.553 & 5.55 &  11.15&   0.360 &2.2   \\
 46374& K2III: & 5.562 & 5.57 &  12.75&   0.598 &2.0   \\
 46758&   G5   & 7.156 & 7.15 &   5.15&   0.411 &2.2   \\
 47138&G8/K0III& 5.704 & 5.71 &  11.79&   0.838 &2.5   \\
 47366& K1III: & 6.117 & 6.11 &  11.75&   1.044 &1.2   \\
 48432& K0III  & 5.349 & 5.34 &  15.69&   0.941 &1.5   \\
 50904&   G5   & 6.539 & 6.53 &   7.02&   0.444 &2.5   \\
 53329&  G8IV  & 5.557 & 5.55 &  10.68&   0.399 &2.0   \\
 54810& K0III  & 4.920 & 4.91 &  15.45&   0.378 &1.5   \\
 55280& K2III  & 5.200 & 5.20 &  16.88&   0.840 &1.5   \\
 56891&   K0   & 6.640 & 6.66 &   7.86&   0.656 &2.0   \\
 58207&G9III+..& 3.793 & 3.78 &  25.9 &   0.453 &2.5   \\
 59686& K2III  & 5.450 & 5.45 &  10.81&   0.123 &2.0   \\
 60294& K2III  & 5.941 & 5.93 &  11.63&   0.717 &1.5   \\
 60318& K0III  & 5.348 & 5.34 &  10.78&   0.189 &3.0   \\
 60341& K0III  & 5.637 & 5.64 &  11.15&   0.364 &2.0   \\
 60986& K0III  & 5.888 & 5.58 &  10.65&   0.744 &2.5   \\
 61363& K0III  & 5.596 & 5.58 &  10.01&   0.184 &1.5   \\
 61935& K0III  & 3.930 & 3.94 &  22.61&   0.284 &2.5   \\
 62141& K0III  & 6.244 & 6.25 &  10.63&   1.058 &2.5   \\
 62345& G8III  & 3.570 & 3.57 &  22.73&   0.109 &3.0   \\
 63798&   G5   & 6.500 & 6.49 &   8.43&   0.825 &2.5   \\
 64152& K0III  & 5.619 & 5.62 &  11.9 &   0.681 &2.5   \\
 64967&  G8IV  & 6.567 & 6.57 &  10.1 &   1.218 &1.2   \\
 65066& K0III  & 6.033 & 6.03 &  11.08&   0.887 &2.0   \\
 65345& K0III  & 5.297 & 5.30 &  12.33&   0.430 &2.5   \\
 67539&   G5   & 6.502 & 6.48 &   5.87& --0.072 &2.0   \\
 68312& G8III  & 5.351 & 5.36 &  10.32&   0.151 &2.5   \\
 68375& G8III  & 5.560 & 5.55 &  11.18&   0.526 &3.0   \\
 70523& K0III  & 5.720 & 5.71 &  11.4 &   0.501 &1.2   \\
 71088& G8III  & 5.889 & 5.89 &  10.12&   0.562 &2.5   \\
 73017&  G8IV  & 5.665 & 5.66 &  13.55&   0.853 &0.9   \\
 74794& K0III: & 5.698 & 5.70 &  11.71&   0.574 &2.0   \\
 75506& K0III  & 5.160 & 5.15 &  11.91&   0.175 &2.0   \\
 75958& G6III  & 5.571 & 5.57 &  10.59&   0.450 &2.7   \\
 76291&  K1IV  & 5.733 & 5.72 &  14.21&   0.891 &0.8   \\
 76813& G9III  & 5.233 & 5.23 &  10.19& --0.006 &3.0   \\
 78235& G8III  & 5.427 & 5.42 &  12.56&   0.646 &2.7   \\
 79181& G8III  & 5.727 & 5.72 &  10.85&   0.535 &1.5   \\
 80546& K3III  & 6.180 & 6.16 &  10.3 &   0.714 &1.5   \\
 81688&K0III-IV& 5.413 & 5.40 &  11.33&   0.272 &1.5   \\
 82969&   G5   & 6.412 & 6.41 &  10.17&   1.119 &1.5   \\
 83240&K1IIIvar& 5.012 & 5.00 &  14.45&   0.333 &2.0   \\
 83371&   K0   & 6.601 & 6.59 &   5.61& --0.027 &2.2   \\
 86513& G9III: & 5.753 & 5.75 &  10.05&   0.331 &2.2   \\
 90633& K2III  & 6.320 & 6.32 &  10.41&   0.873 &1.5   \\
 93291& G4III: & 5.488 & 5.49 &  11.34&   0.481 &3.0   \\
 93875& K2III  & 5.564 & 5.57 &  12.39&   0.491 &1.7   \\
 94084& K2III  & 6.445 & 6.44 &  10.35&   1.106 &2.0   \\
 94402& G8III  & 5.454 & 5.45 &  10.45&   0.246 &3.0   \\
 94497& G7III: & 5.734 & 5.73 &  10.69&   0.413 &1.5   \\
 95808&G7III...& 5.513 & 5.51 &  10.24&   0.234 &3.0   \\
 98366& K0III: & 5.908 & 5.90 &  10.48&   0.544 &2.0   \\
100696& K0III  & 5.203 & 5.19 &  13.49&   0.481 &1.5   \\
101484& K1III  & 5.266 & 5.26 &  14.04&   0.693 &2.5   \\
102928&  K0IV  & 5.635 & 5.62 &  12.59&   0.639 &1.0   \\
103605& K1III  & 5.838 & 5.83 &  10.34&   0.386 &2.0   \\
103912&G7IVw...& 8.390 & 8.35 &   5.55&   1.744 &0.9   \\
104783& G5III  & 9.160 & 9.16 &   2.77&   1.162 &2.2   \\
106714& K0III  & 4.938 & 4.93 &  13.12&   0.192 &3.0   \\
108381&K2IIICN+& 4.350 & 4.35 &  19.18&   0.285 &2.6   \\
109053& G8III  & 9.220 & 9.21 &   --  &   0.595 & --   \\
110024& G9III  & 5.487 & 5.49 &  11.39&   0.427 &2.5   \\
113321   & K0  & 9.388 & 9.35 &   --  &   0.798 & --   \\
113997   & K0  & 9.275 & 9.32 &   --  &   0.489 & --   \\
114357& K3III  & 6.020 & 6.01 &  10.89&   0.640 &1.5   \\
116292& K0III  & 5.367 & 5.36 &  10.2 &   0.068 &3.0   \\
117304& K0III  & 5.656 & 5.65 &  11.99&   0.538 &2.0   \\
117566&G2.5IIIb& 5.750 & 5.74 &  11.15&   0.842 &2.5   \\
119126& G9III  & 5.633 & 5.63 &  10.13&   0.256 &2.5   \\
120084& G7III: & 5.910 & 5.91 &  10.24&   0.600 &2.5   \\
120164&K0III+..& 5.507 & 5.51 &  10.73 &  0.222 &2.5   \\
120420& K0III  & 5.620 & 5.61 &  10.49 &  0.242 &1.5   \\
136138&  G5IV  & 5.694 & 5.68 &  11.23 &  0.638 &2.0   \\
138852&K0III-IV& 5.750 & 5.74 &  10.24 &  0.428 &2.0   \\
139254& K0III  & 5.796 & 5.79 &  11.97 &  0.724 &2.0   \\
139329& K0III  & 5.829 & 5.82 &  11.17 &  0.596 &1.2   \\
143553& K0III: & 6.812 & 5.82 &  13.62 &  1.980 &1.0   \\
146388& K3III  & 5.714 & 5.72 &  10.31 &  0.332 &2.0   \\
148604&G5III/IV& 5.675 & 5.66 &  12.18 &  0.843 &2.5   \\
152224& K0III  & 6.168 & 6.16 &  10.36 &  0.768 &1.0   \\
153956& K1III: & 6.050 & 6.04 &  10.99 &  0.726 &1.5   \\
155970& K1III  & 6.006 & 5.98 &  10.63 &  0.682 &1.5   \\
156874& K0III  & 5.688 & 5.68 &  10.24 &  0.377 &2.5   \\
159353& K0III: & 5.691 & 5.68 &  10.2  &  0.356 &2.5   \\
161178& G9III  & 5.881 & 5.87 &  10.17 &  0.505 &1.5   \\
162076&  G5IV  & 5.700 & 5.69 &  13.04 &  0.952 &2.5   \\
166578&   K0   & 6.683 & 6.67 &   5.83 &  0.138 &1.5   \\
168653& K1III: & 5.969 & 5.96 &  11.48 &  0.758 &1.0   \\
170693&K1,5III & 4.833 & 4.82 &  10.28 &--0.916 &1.0   \\
171994&  G8IV  & 6.316 & 6.31 &  11.14 &  1.251 &1.5   \\
175743& K1III  & 5.706 & 5.69 &  11.84 &  0.586 &2.0   \\
176408& K1III  & 5.676 & 5.67 &  11.4  &  0.405 &1.7   \\
176598& G8III  & 5.632 & 5.62 &  10.39 &  0.420 &3.0   \\
180711& G9III  & 3.082 & 3.07 &  32.54 &  0.252 &2.0   \\
185644& K1III  & 5.302 & 5.30 &  13.31 &  0.386 &1.5   \\
187739& K0III  & 5.905 & 5.88 &  10.46 &  0.503 &1.0   \\
188119& G8III  & 3.830 & 3.84 &  22.4  &  0.272 &2.0   \\
192787& K0III  & 5.711 & 5.70 &  10.86 &  0.578 &2.5   \\
192836& K1III  & 6.127 & 6.11 &  10.95 &  0.901 &2.0   \\
195330&K1/K2III& 6.121 & 6.10 &  10.25 &  0.764 &1.5   \\
196134&K0III-IV& 6.513 & 6.50 &  10.28 &  1.131 &1.7   \\
198431& K1III  & 5.880 & 5.87 &  13.06 &  0.876 &1.0   \\
199870& G8III  & 5.562 & 5.55 &  12.32 &  0.680 &2.5   \\
204771& K0III  & 5.231 & 5.22 &  14.58 &  0.699 &2.5   \\
206005&   K0   & 6.070 & 6.07 &  10.22 &  0.656 &1.2   \\
207130& K0III  & 5.182 & 5.18 &  13.19 &  0.373 &2.5   \\
208111& K2III  & 5.716 & 5.71 &  11.22 &  0.429 &2.0   \\
211006& K2III  & 5.878 & 5.87 &  13.08 &  0.897 &1.5   \\
212496&G8,5IIb & 4.430 & 4.42 &  19.51 &  0.380 &1.0   \\
214567&  G8II  & 5.849 & 5.84 &   8.55 &  0.195 &3.0   \\
215030& G9III  & 5.941 & 5.93 &  10.08 &  0.505 &1.2  \\
215721&  G8II  & 5.260 & 5.24 &  12.26 &  0.345 &1.5  \\
216131& M2III  & 3.513 & 3.51 &  27.95 &  0.432 &2.7  \\
216228& K0III  & 3.510 & 3.50 &  28.27 &  0.298 &1.5  \\
218031& K0IIIb & 4.650 & 4.64 &  18.20 &  0.478 &1.2  \\
219418& G5III  & 6.830 & 6.81 &   5.58 &  0.363 &2.7  \\
219916& K0III  & 4.868 & 4.75 &  15.48 &  0.577 &2.7  \\
221345& K0III  & 5.220 & 5.22 &  13.09 &  0.315 &1.5  \\
221833&   K0   & 6.476 & 6.47 &  10.51 &  1.055 &1.5  \\
225197& K0III  & 5.781 & 5.78 &  11.29 &  0.598 &2.5  \\
225216& K1III  & 5.691 & 5.68 &  10.30 &  0.300 &2.0  \\
BD+222606& K5  & 9.450 & 9.42 &   --   &  1.148 & --  \\
BD+252555&G7III& 9.162 & 9.14 &   --   &  0.594 & --  \\
BD+282250& G5  & 9.370 & 9.34 &   --   &  1.394 & --  \\
\hline
\hline
\end{longtable}

\begin{longtable}{rccccrr}
\caption[]{Parameters of atmospheres of studied stars}\\
\hline
  Star &   \Teff , K & \logg (E)& \logg(Ca) &\Vt, \kms & [Fe/H]$_{I}$& [Fe/H]$_{II}$\\
\hline
\endfirsthead
\caption[]{Continued.}\\
\hline
  Star &   \Teff , K & \logg (E)& \logg(Ca) &\Vt, \kms & [Fe/H]$_{I}$& [Fe/H]$_{II}$]\\
\hline
\endhead
\hline
\endfoot
\label{tableA2}
  2910 &   4756& 2.70 &   2.40 &   1.5 &    0.12 &    0.11 \\
  4188 &   4809& 2.70 &   2.60 &   1.5 &    0.04 &    0.06 \\
  4482 &   4917& 2.65 &   2.65 &   1.4 &    0.02 &    0.03 \\
  5395 &   4849& 2.15 &   2.15 &   1.3 &  --0.32 &  --0.31 \\
  6319 &   4650& 2.30 &   2.30 &   1.3 &    0.06 &    0.07 \\
  6482 &   4738& 2.40 &   2.50 &   1.4 &  --0.11 &  --0.10 \\
  7106 &   4684& 2.55 &   2.45 &   1.5 &    0.05 &    0.06 \\
  7578 &   4680& 2.50 &   2.50 &   1.4 &    0.12 &    0.13 \\
  8207 &   4750& 2.75 &   2.60 &   1.5 &    0.27 &    0.30 \\
  8599 &   4781& 2.50 &   2.40 &   1.1 &  --0.22 &  --0.25 \\
  8733 &   4932& 2.70 &   2.80 &   1.2 &    0.02 &    0.03 \\
  9408 &   4804& 2.30 &   2.40 &   1.5 &  --0.21 &  --0.20 \\
 10975 &   4881& 2.20 &   2.40 &   1.5 &  --0.19 &  --0.16 \\
 11559 &   4977& 3.00 &   2.90 &   1.5 &    0.05 &    0.03 \\
 11749 &   4679& 2.40 &   2.40 &   1.5 &  --0.10 &  --0.06 \\
 11949 &   4708& 2.30 &   2.30 &   1.2 &  --0.16 &  --0.15 \\
 15453 &   4696& 2.40 &   2.40 &   1.3 &  --0.07 &  --0.07 \\
 15755 &   4611& 2.30 &   2.30 &   1.2 &  --0.01 &  --0.02 \\
 15779 &   4821& 2.70 &   2.60 &   1.5 &    0.02 &    0.05 \\
 16247 &   4629& 2.20 &   2.30 &   1.4 &  --0.22 &  --0.20 \\
 16400 &   4840& 2.50 &   2.50 &   1.35&  --0.01 &    0.01 \\
 17361 &   4646& 2.50 &   2.45 &   1.5 &    0.12 &    0.11 \\
 18885 &   4722& 2.50 &   2.60 &   1.4 &    0.16 &    0.16 \\
 19270 &   4723& 2.40 &   2.30 &   1.45&    0.15 &    0.16 \\
 19787 &   4832& 2.75 &   2.65 &   1.5 &    0.14 &    0.15 \\
 19845 &   4933& 2.80 &   2.70 &   1.3 &    0.11 &    0.11 \\
 20791 &   4986& 2.80 &   2.60 &   1.2 &    0.11 &    0.13 \\
 25602 &   4693& 2.40 &   2.40 &   1.15&  --0.42 &  --0.45 \\
 25604 &   4764& 2.70 &   2.60 &   1.5 &    0.13 &    0.14 \\
 26546 &   4743& 2.25 &   2.15 &   1.3 &  --0.01 &  --0.01 \\
 26659 &   5178& 2.90 &   3.00 &   1.2 &  --0.13 &  --0.11 \\
 26755 &   4630& 2.20 &   2.20 &   1.3 &  --0.06 &  --0.03 \\
 27348 &   5003& 2.80 &   2.70 &   1.2 &    0.14 &    0.14 \\
 27371 &   4955& 2.70 &   2.60 &   1.4 &    0.11 &    0.10 \\
 27697 &   4975& 2.65 &   2.55 &   1.4 &    0.11 &    0.07 \\
 28292 &   4453& 2.10 &   2.10 &   1.5 &  --0.18 &  --0.17 \\
 28305 &   4925& 2.55 &   2.45 &   1.4 &    0.11 &    0.11 \\
 28307 &   4961& 2.70 &   2.75 &   1.3 &    0.12 &    0.08 \\
 30557 &   4829& 2.45 &   2.45 &   1.35&  --0.07 &  --0.05 \\
 31444 &   5080& 2.75 &   2.75 &   1.2 &  --0.17 &  --0.13 \\
 33419 &   4708& 2.30 &   2.30 &    1.4 &    0.00 &    0.05  \\
 33618 &   4590& 2.30 &   2.25 &    1.4 &    0.05 &    0.05  \\
 34200 &   5055& 2.80 &   2.80 &    1.3 &    0.04 &    0.06  \\
 34559 &   5010& 2.90 &   2.90 &    1.2 &    0.04 &    0.06  \\
 35369 &   4931& 2.40 &   2.40 &    1.4 &  --0.14 &  --0.14  \\
 37638 &   5093& 2.80 &   2.80 &    1.3 &  --0.01 &    0.01  \\
 39070 &   5047& 2.80 &   2.80 &    1.15&    0.03 &    0.05  \\
 39910 &   4618& 2.60 &   2.60 &    1.35&    0.27 &    0.26  \\
 40020 &   4670& 2.30 &   2.40 &    1.5 &    0.13 &    0.14  \\
 40801 &   4703& 2.20 &   2.20 &    1.05&  --0.21 &  --0.23  \\
 42341 &   4655& 2.60 &   2.80 &    1.4 &    0.25 &    0.22  \\
 43023 &   4994& 2.40 &   2.40 &    1.3 &  --0.13 &  --0.12  \\
 45415 &   4762& 2.30 &   2.30 &    1.3 &    0.03 &    0.02  \\
 46374 &   4661& 2.30 &   2.30 &    1.5 &    0.03 &    0.03  \\
 46758 &   5003& 2.90 &   2.80 &    1.3 &  --0.30 &  --0.32  \\
 47138 &   5211& 3.00 &   3.00 &    1.2 &  --0.06 &  --0.04  \\
 47366 &   4772& 2.60 &   2.60 &    1.2 &  --0.16 &  --0.13  \\
 48432 &   4836& 2.65 &   2.65 &    1.3 &  --0.29 &  --0.31  \\
 50904 &   4953& 2.70 &   2.70 &    1.3 &  --0.13 &  --0.08  \\
 53329 &   5012& 2.80 &   2.80 &    1.2 &  --0.38 &  --0.40  \\
 54810 &   4669& 2.40 &   2.40 &    1.4 &  --0.47 &  --0.49  \\
 55280 &   4654& 2.25 &   2.25 &    1.3 &  --0.08 &  --0.07  \\
 56891 &   4709& 2.40 &   2.40 &    1.4 &    0.11 &    0.07  \\
 58207 &   4799& 2.35 &   2.35 &    1.4 &  --0.14 &  --0.09  \\
 59686 &   4654& 2.40 &   2.40 &    1.4 &    0.02 &    0.03  \\
 60294 &   4569& 2.15 &   2.15 &    1.3 &  --0.08 &  --0.07  \\
 60318 &   4962& 2.80 &   2.80 &    1.2 &    0.08 &    0.11  \\
 60341 &   4634& 2.15 &   2.15 &    1.4 &  --0.02 &    0.01  \\
 60986 &   5057& 2.60 &   2.60 &    1.3 &  --0.01 &    0.00  \\
 61363 &   4785& 2.10 &   2.10 &    1.1 &  --0.21 &  --0.17  \\
 61935 &   4780& 2.40 &   2.30 &    1.3 &  --0.09 &  --0.09  \\
 62141 &   4971& 2.80 &   2.70 &    1.3 &  --0.14 &  --0.13  \\
 62345 &   5032& 2.60 &   2.50 &    1.2 &    0.05 &    0.09  \\
 63798 &   5004& 2.50 &   2.50 &    1.3 &  --0.10 &  --0.08  \\
 64152 &   4977& 2.70 &   2.70 &    1.3 &  --0.01 &  --0.01  \\
 64967 &   4864& 2.55 &   2.55 &    1.5 &  --0.65 &  --0.64  \\
 65066 &   4868& 2.60 &   2.60 &    1.5 &    0.02 &    0.04  \\
 65345 &   4963& 2.70 &   2.60 &    1.3 &    0.00 &    0.03  \\
 67539 &   4781& 2.45 &   2.45 &    1.2 &  --0.61 &  --0.63  \\
 68312 &   5090& 2.70 &   2.70 &    1.3 &  --0.09 &  --0.10  \\
 68375 &   5071& 2.90 &   2.90 &    1.3 &    0.00 &    0.02 \\
 70523 &   4642& 2.20 &   2.10 &    1.4 &  --0.25 &  --0.26 \\
 71088 &   4900& 2.70 &   2.70 &    1.3 &  --0.03 &  --0.01 \\
 73017 &   4693& 2.30 &   2.40 &    1.2 &  --0.66 &  --0.64 \\
 74794 &   4701& 2.25 &   2.25 &    1.4 &  --0.02 &  --0.01 \\
 75506 &   4876& 2.50 &   2.30 &    1.3 &  --0.30 &  --0.31 \\
 75958 &   5030& 2.70 &   2.80 &    1.3 &  --0.09 &  --0.07 \\
 76291 &   4495& 2.00 &   2.20 &    1.3 &  --0.28 &  --0.30 \\
 76813 &   5060& 2.80 &   2.80 &    1.4 &  --0.09 &  --0.07 \\
 78235 &   5070& 2.80 &   2.80 &    1.3 &  --0.14 &  --0.14 \\
 79181 &   4867& 2.40 &   2.40 &    1.2 &  --0.28 &  --0.25 \\
 80546 &   4601& 2.25 &   2.35 &    1.3 &  --0.05 &  --0.03 \\
 81688 &   4789& 2.30 &   2.30 &    1.3 &  --0.23 &  --0.21 \\
 82969 &   4948& 2.70 &   3.00 &    1.2 &  --0.22 &  --0.20 \\
 83240 &   4682& 2.45 &   2.45 &    1.3 &  --0.02 &    0.01 \\
 83371 &   4861& 2.60 &   2.50 &    1.3 &  --0.39 &  --0.39 \\
 86513 &   4755& 2.30 &   2.30 &    1.4 &  --0.08 &    0.07 \\
 90633 &   4596& 2.30 &   2.30 &    1.3 &    0.02 &    0.03 \\
 93291 &   5061& 2.75 &   2.75 &    1.3 &  --0.05 &    0.05 \\
 93875 &   4590& 2.25 &   2.25 &    1.4 &    0.06 &    0.01 \\
 94084 &   4787& 2.65 &   2.45 &    1.4 &    0.11 &    0.11 \\
 94402 &   5004& 2.70 &   2.70 &    1.4 &    0.11 &    0.13 \\
 94497 &   4702& 2.30 &   2.30 &    1.3 &  --0.19 &  --0.18 \\
 95808 &   4946& 2.55 &   2.55 &    1.4 &  --0.09 &  --0.11 \\
 98366 &   4702& 2.40 &   2.40 &    1.2 &  --0.10 &  --0.11 \\
100696 &   4862& 2.40 &   2.30 &    1.4 &  --0.31 &  --0.31 \\
101484 &   4991& 2.70 &   2.70 &    1.3 &  --0.03 &  --0.03 \\
102928 &   4654& 2.35 &   2.25 &    1.4 &  --0.28 &  --0.28 \\
103605 &   4611& 2.35 &   2.35 &    1.4 &  --0.10 &  --0.08 \\
103912 &   4870& 2.80 &   2.80 &    1.1 &  --0.65 &  --0.68 \\
104783 &   5247& 2.55 &   2.55 &    1.5 &  --0.36 &  --0.34 \\
106714 &   4935& 2.50 &   2.50 &    1.2 &  --0.09 &  --0.05 \\
108381 &   4680& 2.50 &   2.50 &    1.3 &    0.21 &    0.17 \\
109053 &   4921& 2.50 &    --  &    1.4 &  --0.38 &   --    \\
110024 &   4921& 2.70 &   2.70 &    1.4 &    0.06 &    0.08 \\
113321 &   4739& 2.10 &    --  &    1.4 &  --0.07 &   --    \\
113997 &   4697& 1.70 &    --  &    1.5 &  --0.12 &   --    \\
114357 &   4551& 2.30 &   2.50 &    1.5 &    0.12 &    0.11 \\
116292 &   4922& 2.60 &   2.60 &    1.5 &  --0.03 &  --0.01 \\
117304 &   4630& 2.17 &   2.35 &    1.3 &  --0.15 &  --0.12 \\
117566 &   5475& 3.15 &   3.15 &    1.35&    0.09 &    0.08 \\
119126 &   4802& 2.25 &   2.25 &    1.35&  --0.12 &  --0.11 \\
120084 &   4883& 2.55 &   2.65 &    1.5 &    0.09 &    0.08 \\
120164 &   4746& 2.30 &   2.30 &    1.5 &  --0.07 &  --0.08 \\
120420 &   4676& 2.15 &   2.30 &    1.25&  --0.27 &  --0.24 \\
136138 &   4995& 2.60 &   2.80 &    1.5 &  --0.19 &  --0.19 \\
138852 &   4859& 2.30 &   2.30 &    1.4 &  --0.24 &  --0.21 \\
139254 &   4708& 2.35 &   2.35 &    1.4 &  --0.04 &  --0.02 \\
139329 &   4690& 2.30 &   2.20 &    1.4 &  --0.31 &  --0.33 \\
143553 &   4644& 2.30 &   2.15 &    1.0 &  --0.36 &  --0.36 \\
146388 &   4731& 2.45 &   2.45 &    1.4 &    0.08 &    0.08 \\
148604 &   5110& 2.80 &   2.80 &    1.1 &  --0.18 &  --0.16 \\
152224 &   4685& 2.25 &   2.15 &    1.4 &  --0.24 &  --0.25 \\
153956 &   4604& 2.45 &   2.45 &    1.5 &    0.25 &    0.24 \\
155970 &   4717& 2.50 &   2.50 &    1.3 &    0.08 &    0.09 \\
156874 &   4881& 2.50 &   2.50 &    1.3 &    0.00 &    0.02 \\
159353 &   4850& 2.40 &   2.40 &    1.35&  --0.08 &  --0.08 \\
161178 &   4789& 2.20 &   2.40 &    1.3 &  --0.24 &  --0.24 \\
162076 &   4959& 2.70 &   2.70 &    1.3 &  --0.03 &    0.01 \\
166578 &   4859& 2.50 &   2.55 &    1.4 &  --0.62 &  --0.64 \\
168653 &   4632& 2.20 &   2.40 &    1.4 &  --0.16 &  --0.16 \\
170693 &   4256& 1.50 &   1.50 &    1.25&  --0.59 &  --0.61 \\
171994 &   5014& 2.70 &   2.70 &    1.2 &  --0.23 &  --0.22 \\
175743 &   4669& 2.50 &   2.40 &    1.4 &    0.04 &    0.04 \\
176408 &   4564& 2.25 &   2.25 &    1.5 &    0.04 &    0.03 \\
176598 &   5024& 2.80 &   2.80 &    1.3 &    0.03 &    0.04 \\
180711 &   4824& 2.40 &   2.40 &    1.3 &  --0.20 &  --0.17 \\
185644 &   4591& 2.40 &   2.40 &    1.20&    0.01 &  --0.01 \\
187739 &   4649& 2.30 &   2.10 &    1.2 &  --0.34 &  --0.37 \\
188119 &   4993& 2.75 &   2.60 &    1.1 &  --0.31 &  --0.28 \\
192787 &   4987& 2.60 &   2.60 &    1.35&  --0.05 &  --0.97 \\
192836 &   4772& 2.60 &   2.55 &    1.35&    0.01 &    0.00 \\
195330 &   4792& 2.40 &   2.50 &    0.6 &  --0.29 &  --0.30 \\
196134 &   4741& 2.40 &   2.50 &    1.3 &  --0.14 &  --0.12 \\
198431 &   4524& 2.00 &   2.10 &    1.2 &  --0.37 &  --0.39 \\
199870 &   4937& 2.70 &   2.70 &    1.3 &    0.02 &  --0.01 \\
204771 &   4904& 2.70 &   2.70 &    1.2 &    0.02 &  --0.01 \\
206005 &   4709& 2.20 &   2.20 &    1.2 &  --0.19 &  --0.19 \\
207130 &   4792& 2.60 &   2.50 &    1.2 &    0.13 &    0.12 \\
208111 &   4592& 2.30 &   2.10 &    1.5 &    0.18 &    0.21 \\
211006 &   4553& 2.35 &   2.35 &    1.4 &    0.07 &    0.05 \\
212496 &   4646& 2.30 &   2.30 &    1.2 &  --0.48 &  --0.49 \\
214567 &   4981& 2.50 &   2.50 &    1.3 &  --0.14 &  --0.13 \\
215030 &   4723& 2.35 &   2.35 &    1.25&  --0.49 &  --0.49 \\
215721 &   4890& 2.40 &   2.40 &    1.3 &  --0.50 &  --0.52 \\
216131 &   4984& 2.70 &   2.70 &    1.2 &  --0.07 &  --0.07 \\
216228 &   4698& 2.40 &   2.40 &    1.4 &  --0.16 &  --0.13 \\
218031 &   4692& 2.20 &   2.30 &    1.4 &  --0.24 &  --0.22 \\
219418 &   5281& 2.80 &   2.90 &    1.3 &  --0.27 &  --0.29 \\
219916 &   5038& 2.80 &   2.80 &    1.2 &    0.03 &    0.01 \\
221345 &   4664& 2.20 &   2.10 &    1.4 &  --0.37 &  --0.41 \\
221833 &   4603& 2.30 &   2.30 &    1.4 &    0.02 &    0.03 \\
225197 &   4734& 2.50 &   2.40 &    1.4 &    0.14 &    0.13 \\
225216 &   4720& 2.20 &   2.20 &    1.4 &  --0.15 &  --0.13 \\
BD+22 2606 &4680 &2.10 &--  &        1.3&    --0.26  &   --  \\
BD+25 2555 &5014 &2.80 &--  &        1.5&    --0.43  &   --  \\
BD+28 2250 &4630 &2.00 &--  &        1.2&    --0.69  &   --  \\
\hline
\hline
\end{longtable}

\begin{longtable}{rrrrrrrrr}
\caption[]{Si, Ca, Ni abundances}\\
\hline
HD  & [Fe/H]& $\sigma$&[Si/Fe]& $\sigma$&[Ca/Fe]&$\sigma$&[Ni/Fe]&$\sigma$\\
\hline
\endfirsthead
\caption[]{Continued.}\\
\hline
HD  & [Fe/H]& $\sigma$&[Si/Fe]& $\sigma$&[Ca/Fe]&$\sigma$&[Ni/Fe]&$\sigma$\\
\hline
\endhead
\hline
\endfoot
\label{tableA3}
2910  &   0.12 &   0.11 &     0.16 &   0.10 &    0.02&    0.16&     0.05&    0.13 \\
4188  &   0.04 &   0.13 &     0.16 &   0.13 &    0.02&    0.07&     0.01&    0.16 \\
4482  &   0.02 &   0.10 &     0.12 &   0.08 &    0.05&    0.06&   --0.10&    0.09 \\
5395  & --0.32 &   0.06 &     0.13 &   0.10 &    0.05&    0.08&   --0.02&    0.06 \\
6319  &   0.06 &   0.09 &     0.11 &   0.09 &    0.02&    0.09&     0.01&    0.08 \\
6482  & --0.11 &   0.07 &     0.10 &   0.07 &    0.07&    0.10&   --0.02&    0.09 \\
7106  &   0.05 &   0.11 &     0.18 &   0.10 &    0.05&    0.12&     0.08&    0.12 \\
7578  &   0.12 &   0.13 &     0.14 &   0.05 &    0.03&    0.14&     0.12&    0.12 \\
8207  &   0.27 &   0.12 &     0.17 &   0.11 &    0.03&    0.15&     0.05&    0.14 \\
8599  & --0.22 &   0.05 &     0.09 &   0.05 &    0.10&    0.13&   --0.01&    0.07 \\
8733  &   0.02 &   0.06 &     0.03 &   0.08 &    0.06&    0.09&   --0.02&    0.09 \\
9408  & --0.21 &   0.10 &     0.15 &   0.08 &    0.03&    0.07&     0.06&    0.08 \\
10975 & --0.19 &   0.12 &     0.10 &   0.10 &  --0.01&    0.06&   --0.03&    0.12 \\
11559 &   0.05 &   0.14 &     0.15 &   0.16 &    0.06&    0.11&   --0.07&    0.14 \\
11749 & --0.10 &   0.15 &     0.20 &   0.15 &    0.00&    0.11&     0.00&    0.17 \\
11949 & --0.16 &   0.07 &     0.12 &   0.06 &    0.14&    0.08&     0.00&    0.09 \\
15453 & --0.07 &   0.13 &     0.14 &   0.09 &    0.01&    0.02&     0.00&    0.03 \\
15755 & --0.01 &   0.09 &     0.14 &   0.09 &  --0.02&    0.11&     0.03&    0.08 \\
15779 &   0.02 &   0.13 &     0.13 &   0.12 &    0.02&    0.07&   --0.05&    0.16 \\
16247 & --0.22 &   0.16 &     0.18 &   0.08 &    0.00&    0.14&     0.00&    0.13 \\
16400 & --0.01 &   0.08 &     0.05 &   0.07 &    0.05&    0.10&     0.01&    0.12 \\
17361 &   0.12 &   0.13 &     0.18 &   0.10 &  --0.01&    0.18&     0.07&    0.15 \\
18885 &   0.16 &   0.09 &     0.04 &   0.04 &  --0.07&    0.09&     0.01&    0.11 \\
19270 &   0.15 &   0.16 &     0.15 &   0.10 &    0.04&    0.12&     0.12&    0.09 \\
19787 &   0.14 &   0.13 &     0.17 &   0.13 &    0.07&    0.11&   --0.05&    0.16 \\
19845 &   0.11 &   0.09 &     0.17 &   0.10 &    0.02&    0.07&     0.10&    0.12 \\
20791 &   0.11 &   0.08 &     0.13 &   0.12 &  --0.02&    0.10&     0.02&    0.12 \\
25602 & --0.42 &   0.12 &     0.22 &   0.03 &    0.15&    0.11&     0.04&    0.08 \\
25604 &   0.13 &   0.14 &     0.23 &   0.09 &  --0.03&    0.21&     0.04&    0.19 \\
26546 & --0.01 &   0.12 &     0.10 &   0.12 &    0.14&    0.13&     0.02&    0.15 \\
26659 & --0.13 &   0.09 &     0.03 &   0.09 &    0.06&    0.12&   --0.04&    0.06 \\
26755 & --0.06 &   0.11 &     0.14 &   0.10 &    0.08&    0.13&   --0.03&    0.11 \\
27348 &   0.14 &   0.11 &     0.12 &   0.11 &  --0.06&    0.10&     0.14&    0.08 \\
27371 &   0.11 &   0.10 &     0.07 &   0.12 &    0.10&    0.12&   --0.04&    0.12 \\
27697 &   0.11 &   0.09 &     0.07 &   0.11 &    0.08&    0.12&     0.06&    0.09 \\
28292 & --0.18 &   0.16 &     0.24 &   0.11 &    0.01&    0.10&   --0.06&    0.13 \\
28305 &   0.11 &   0.09 &     0.09 &   0.11 &    0.11&    0.12&     0.09&    0.11 \\
28307 &   0.12 &   0.13 &     0.06 &   0.10 &    0.04&    0.12&     0.00&    0.12 \\
30557 & --0.07 &   0.11 &     0.07 &   0.08 &    0.01&    0.09&     0.02&    0.13 \\
31444 & --0.17 &   0.10 &     0.06 &   0.13 &    0.14&    0.13&   --0.07&    0.08 \\
33419 &   0.00 &   0.10 &     0.12 &   0.09 &    0.08&    0.06&     0.12&    0.10 \\
33618 &   0.05 &   0.11 &     0.14 &   0.09 &    0.04&    0.12&     0.08&    0.13 \\
34200 &   0.04 &   0.08 &     0.04 &   0.07 &    0.00&    0.06&   --0.07&    0.04 \\
34559 &   0.04 &   0.09 &     0.06 &   0.10 &    0.05&    0.11&   --0.09&    0.08 \\
35369 & --0.14 &   0.12 &     0.05 &   0.10 &  --0.02&    0.08&   --0.02&    0.11 \\
37638 & --0.01 &   0.12 &     0.06 &   0.09 &    0.01&    0.02&   --0.07&    0.06 \\
39070 &   0.03 &   0.07 &     0.01 &   0.07 &  --0.02&    0.01&   --0.03&    0.06 \\
39910 &   0.27 &   0.10 &     0.09 &   0.13 &  --0.07&    0.05&     0.15&    0.10 \\
40020 &   0.13 &   0.11 &     0.15 &   0.14 &  --0.05&    0.02&     0.05&    0.11 \\
40801 & --0.21 &   0.07 &     0.07 &   0.06 &    0.11&    0.09&     0.03&    0.07 \\
42341 &   0.25 &   0.12 &     0.16 &   0.14 &  --0.01&    0.01&     0.12&    0.10 \\
43023 & --0.13 &   0.12 &     0.04 &   0.06 &    0.11&    0.10&   --0.07&    0.08 \\
45415 &   0.03 &   0.12 &   --0.01 &    --  &  --0.07&    0.10&     0.03&    0.12 \\
46374 &   0.03 &   0.13 &     0.10 &   0.09 &  --0.02&    0.12&   --0.05&    0.14 \\
46758 & --0.30 &   0.09 &     0.14 &   0.10 &    0.06&    0.10&   --0.01&    0.09 \\
47138 & --0.06 &   0.06 &     0.03 &   0.09 &    0.05&    0.07&   --0.06&    0.07 \\
47366 & --0.16 &   0.11 &     0.12 &   0.07 &    0.06&    0.09&     0.00&    0.09 \\
48432 & --0.29 &   0.10 &     0.17 &   0.08 &    0.14&    0.08&   --0.04&    0.10 \\
50904 & --0.13 &   0.09 &     0.13 &   0.10 &    0.07&    0.10&   --0.06&    0.10 \\
53329 & --0.38 &   0.10 &     0.08 &   0.10 &    0.05&    0.06&   --0.04&    0.07 \\
54810 & --0.47 &   0.13 &     0.22 &   0.11 &    0.11&    0.07&   --0.02&    0.07 \\
55280 & --0.08 &   0.12 &     0.13 &   0.09 &    0.07&    0.06&     0.01&    0.12 \\
56891 &   0.11 &   0.11 &     0.11 &   0.14 &    0.01&    0.07&     0.10&    0.12 \\
58207 & --0.14 &   0.11 &     0.12 &   0.09 &    0.09&    0.08&     0.03&    0.10 \\
59686 &   0.02 &   0.11 &     0.19 &   0.09 &    0.03&    0.11&     0.06&    0.16 \\
60294 & --0.08 &   0.13 &     0.16 &   0.11 &    0.03&    0.05&     0.02&    0.11 \\
60318 &   0.08 &   0.11 &     0.10 &   0.11 &  --0.01&    0.07&     0.08&    0.12 \\
60341 & --0.02 &   0.11 &     0.10 &   0.08 &    0.05&    0.14&     0.04&    0.13 \\
60986 & --0.01 &   0.10 &     0.10 &   0.10 &    0.10&    0.09&   --0.03&    0.08 \\
61363 & --0.21 &   0.30 &     0.04 &   0.07 &    0.01&    0.08&     0.01&    0.11 \\
61935 & --0.09 &   0.13 &     0.10 &   0.07 &    0.12&    0.13&     0.01&    0.07 \\
62141 & --0.14 &   0.11 &     0.14 &   0.12 &  --0.03&    0.12&   --0.09&    0.10 \\
62345 &   0.05 &   0.11 &     0.05 &   0.11 &    0.08&    0.07&   --0.06&    0.11 \\
63798 & --0.10 &   0.11 &     0.09 &   0.10 &    0.09&    0.07&   --0.03&    0.11 \\
64152 & --0.01 &   0.10 &     0.09 &   0.09 &    0.07&    0.07&     0.00&    0.07 \\
64967 & --0.65 &   0.13 &     0.34 &   0.07 &    0.14&    0.12&   --0.02&    0.12 \\
65066 &   0.02 &   0.11 &     0.11 &   0.09 &    0.06&    0.10&     0.00&    0.12 \\
65345 &   0.00 &   0.11 &     0.12 &   0.10 &  --0.04&    0.08&     0.00&    0.11 \\
67539 & --0.61 &   0.08 &     0.25 &   0.11 &    0.23&    0.10&     0.06&    0.06 \\
68312 & --0.09 &   0.12 &     0.06 &   0.10 &    0.06&    0.10&   --0.07&    0.10 \\
68375 &   0.00 &   0.11 &     0.07 &   0.10 &  --0.06&    0.05&   --0.03&    0.10 \\
70523 & --0.25 &   0.11 &     0.24 &   0.08 &    0.10&    0.12&     0.03&    0.11 \\
71088 & --0.03 &   0.12 &     0.07 &   0.06 &  --0.06&    0.06&   --0.03&    0.11 \\
73017 & --0.66 &   0.10 &     0.21 &   0.09 &    0.02&    0.04&   --0.01&    0.08 \\
74794 & --0.02 &   0.12 &     0.09 &   0.12 &    0.06&    0.09&     0.04&    0.15 \\
75506 & --0.30 &   0.10 &     0.07 &   0.10 &    0.10&    0.10&     0.02&    0.11 \\
75958 & --0.09 &   0.12 &     0.09 &   0.10 &  --0.06&    0.02&   --0.10&    0.08 \\
76291 & --0.28 &   0.10 &     0.21 &   0.11 &    0.08&    0.10&   --0.01&    0.08 \\
76813 & --0.09 &   0.08 &     0.10 &   0.07 &  --0.03&    0.05&   --0.08&    0.10 \\
78235 & --0.14 &   0.11 &     0.11 &   0.10 &    0.07&    0.11&   --0.09&    0.09 \\
79181 & --0.28 &   0.09 &     0.14 &   0.08 &    0.11&    0.13&     0.05&    0.10 \\
80546 & --0.05 &   0.13 &     0.12 &   0.09 &    0.11&    0.14&     0.02&    0.10 \\
81688 & --0.23 &   0.10 &     0.10 &   0.09 &  --0.01&    0.07&   --0.04&    0.09 \\
82969 & --0.22 &   0.09 &     0.07 &   0.10 &    0.00&    0.09&   --0.11&    0.09 \\
83240 & --0.02 &   0.13 &     0.08 &   0.10 &  --0.03&    0.09&   --0.02&    0.12 \\
83371 & --0.39 &   0.07 &     0.13 &   0.11 &    0.08&    0.12&   --0.03&    0.11 \\
86513 & --0.08 &   0.13 &     0.10 &   0.12 &    0.12&    0.08&   --0.02&    0.14 \\
90633 &   0.02 &   0.12 &     0.14 &   0.09 &  --0.01&    0.10&   --0.02&    0.09 \\
93291 & --0.05 &   0.11 &     0.00 &   0.07 &    0.05&    0.12&   --0.06&    0.06 \\
93875 &   0.06 &   0.10 &     0.15 &   0.14 &    0.02&    0.10&     0.05&    0.10 \\
94084 &   0.11 &   0.10 &     0.16 &   0.11 &    0.01&    0.11&     0.04&    0.11 \\
94402 &   0.11 &   0.10 &     0.05 &   0.09 &    0.04&    0.14&     0.06&    0.16 \\
94497 & --0.19 &   0.13 &     0.15 &   0.10 &    0.00&    0.11&   --0.03&    0.09 \\
95808 & --0.09 &   0.12 &     0.11 &   0.07 &    0.07&    0.04&   --0.03&    0.12 \\
98366 & --0.1  &   0.12 &     0.09 &   0.07 &    0.06&    0.11&     0.00&    0.11 \\
100696& --0.31 &   0.13 &     0.11 &   0.09 &    0.07&    0.04&   --0.06&    0.09 \\
101484& --0.03 &   0.12 &     0.13 &   0.06 &    0.05&    0.09&     0.02&    0.11 \\
102928& --0.28 &   0.12 &     0.21 &   0.13 &    0.11&    0.11&   --0.04&    0.14 \\
103605& --0.10 &   0.12 &     0.19 &   0.06 &    0.01&    0.11&     0.00&    0.13 \\
103912& --0.65 &   0.12 &     0.28 &   0.10 &    0.19&    0.08&   --0.03&    0.11 \\
104783& --0.36 &   0.10 &     0.19 &   0.08 &    0.10&    0.12&     0.06&    0.09 \\
106714& --0.09 &   0.10 &     0.00 &   0.06 &    0.01&    0.08&   --0.04&    0.12 \\
108381&   0.21 &   0.09 &     0.17 &   0.13 &    0.01&    0.11&     0.15&    0.14 \\
109053& --0.38 &   0.14 &     0.10 &   0.08 &    0.08&    0.12&   --0.03&    0.09 \\
110024&   0.06 &   0.10 &     0.12 &   0.08 &    0.00&    0.08&     0.02&    0.12 \\
113321& --0.07 &   0.12 &     0.09 &   0.12 &    0.04&    0.10&     0.04&    0.13 \\
113997& --0.12 &   0.15 &     0.08 &   0.14 &    0.18&    0.19&   --0.02&    0.12 \\
114357&   0.12 &   0.13 &     0.15 &   0.11 &  --0.08&    0.10&     0.11&    0.14 \\
116292& --0.03 &   0.11 &     0.08 &   0.07 &    0.00&    0.07&   --0.06&    0.12 \\
117304& --0.15 &   0.12 &     0.15 &   0.13 &  --0.03&    0.09&     0.01&    0.10 \\
117566&   0.09 &   0.10 &   --0.05 &   0.10 &  --0.06&    0.14&   --0.05&    0.08 \\
119126& --0.12 &   0.11 &     0.12 &   0.10 &    0.08&    0.12&     0.04&    0.08 \\
120084&   0.09 &   0.10 &     0.08 &   0.10 &  --0.08&    0.10&     0.00&    0.09 \\
120164& --0.07 &   0.13 &     0.13 &   0.08 &  --0.05&    0.04&   --0.04&    0.13 \\
120420& --0.27 &   0.12 &     0.10 &   0.07 &    0.07&    0.07&   --0.02&    0.10 \\
136138& --0.19 &   0.13 &     0.14 &   0.09 &    0.06&    0.07&   --0.04&    0.09 \\
138852& --0.24 &   0.12 &     0.09 &   0.09 &    0.04&    0.05&   --0.03&    0.08 \\
139254& --0.04 &   0.13 &     0.10 &   0.09 &  --0.04&    0.12&   --0.05&    0.13 \\
139329& --0.31 &   0.12 &     0.15 &   0.08 &    0.10&    0.07&   --0.05&    0.10 \\
143553& --0.36 &   0.10 &     0.16 &   0.10 &    0.14&    0.08&     0.05&    0.10 \\
146388&   0.08 &   0.12 &     0.10 &   0.09 &    0.12&    0.07&     0.09&    0.08 \\
148604& --0.18 &   0.11 &     0.08 &   0.10 &    0.04&    0.10&   --0.07&    0.08 \\
152224& --0.24 &   0.15 &     0.08 &   0.07 &    0.05&    0.11&   --0.06&    0.09 \\
153956&   0.25 &   0.13 &     0.11 &   0.11 &  --0.10&    0.10&     0.05&    0.14 \\
155970&   0.08 &   0.10 &     0.17 &   0.08 &    0.01&    0.10&     0.08&    0.11 \\
156874&   0.00 &   0.12 &     0.04 &   0.13 &  --0.07&    0.03&   --0.06&    0.12 \\
159353& --0.08 &   0.11 &     0.13 &   0.08 &    0.11&    0.09&     0.02&    0.09 \\
161178& --0.24 &   0.09 &     0.06 &   0.10 &    0.05&    0.08&     0.01&    0.08 \\
162076& --0.03 &   0.11 &     0.11 &   0.09 &    0.08&    0.08&   --0.05&    0.09 \\
166578& --0.62 &   0.09 &     0.24 &   0.09 &    0.16&    0.05&   --0.01&    0.08 \\
168653& --0.16 &   0.13 &     0.16 &   0.11 &  --0.01&    0.09&     0.00&    0.14 \\
170693& --0.59 &   0.08 &     0.26 &   0.12 &  --0.02&    0.06&     0.01&    0.11 \\
171994& --0.23 &   0.10 &     0.08 &   0.08 &    0.05&    0.04&   --0.10&    0.09 \\
175743&   0.04 &   0.13 &     0.12 &   0.08 &  --0.01&    0.12&     0.06&    0.15 \\
176408&   0.04 &   0.14 &     0.17 &   0.11 &    0.05&    0.14&     0.02&    0.15 \\
176598&   0.03 &   0.10 &     0.05 &   0.06 &    0.00&    0.12&   --0.08&    0.10 \\
180711& --0.20 &   0.11 &     0.08 &   0.08 &    0.06&    0.10&     0.03&    0.08 \\
185644&   0.01 &   0.11 &     0.07 &   0.05 &  --0.08&    0.13&   --0.04&    0.09 \\
187739& --0.34 &   0.13 &     0.14 &   0.08 &    0.06&    0.12&   --0.08&    0.12 \\
188119& --0.31 &   0.10 &     0.12 &   0.10 &    0.02&    0.10&     0.03&    0.13 \\
192787& --0.05 &   0.12 &     0.04 &   0.06 &  --0.05&    0.07&   --0.07&    0.12 \\
192836&   0.01 &   0.12 &     0.10 &   0.08 &    0.05&    0.11&     0.11&    0.06 \\
195330& --0.29 &   0.11 &     0.11 &   0.12 &    0.11&    0.15&     0.00&    0.13 \\
196134& --0.14 &   0.09 &     0.11 &   0.07 &    0.00&    0.12&     0.02&    0.10 \\
198431& --0.37 &   0.09 &     0.18 &   0.10 &    0.04&    0.11&   --0.03&    0.10 \\
199870&   0.02 &   0.13 &     0.16 &   0.09 &    0.10&    0.14&     0.09&    0.14 \\
204771&   0.02 &   0.10 &     0.05 &   0.07 &    0.06&    0.12&   --0.01&    0.07 \\
206005& --0.19 &   0.11 &     0.07 &   0.06 &    0.16&    0.15&     0.00&    0.11 \\
207130&   0.13 &   0.11 &     0.02 &   0.08 &    0.01&    0.12&     0.09&    0.14 \\
208111&   0.18 &   0.11 &     0.12 &   0.06 &  --0.09&    0.13&     0.03&    0.16 \\
211006&   0.07 &   0.10 &     0.22 &   0.09 &    0.05&    0.12&     0.07&    0.14 \\
212496& --0.48 &   0.09 &     0.17 &   0.08 &    0.12&    0.08&   --0.02&    0.05 \\
214567& --0.14 &   0.12 &     0.08 &   0.10 &  --0.01&    0.04&   --0.06&    0.11 \\
215030& --0.49 &   0.12 &     0.21 &   0.10 &    0.12&    0.10&     0.01&    0.07 \\
215721& --0.50 &   0.09 &     0.12 &   0.09 &    0.15&    0.10&     0.01&    0.08 \\
216131& --0.07 &   0.13 &     0.06 &   0.10 &    0.04&    0.10&   --0.02&    0.12 \\
216228& --0.16 &   0.14 &     0.16 &   0.10 &    0.08&    0.11&     0.01&    0.13 \\
218031& --0.24 &   0.09 &     0.15 &   0.11 &  --0.02&    0.08&   --0.03&    0.12 \\
219418& --0.27 &   0.09 &     0.07 &   0.10 &    0.02&    0.09&   --0.02&    0.05 \\
219916&   0.03 &   0.11 &     0.02 &   0.07 &  --0.03&    0.06&   --0.06&    0.12 \\
221345& --0.37 &   0.13 &     0.20 &   0.10 &    0.07&    0.04&     0.03&    0.13 \\
221833&   0.02 &   0.12 &     0.11 &   0.08 &    0.06&    0.13&     0.01&    0.16 \\
225197&   0.14 &   0.13 &     0.03 &   0.12 &    0.00&    0.10&     0.10&    0.16 \\
225216& --0.15 &   0.10 &     0.11 &   0.06 &    0.09&    0.13&     0.05&    0.10 \\
BD+222606 &--0.26 & 0.12  &0.10 &  0.07&    0.00 &   0.05 &   0.02&    0.12 \\
BD+252555 &--0.43 & 0.10  &0.22 &  0.05&    0.17 &   0.09 & --0.02&    0.11 \\
BD+282250 &--0.69 & 0.13  &0.29 &  0.11&    0.38 &   0.09 &   0.00&    0.14 \\
\hline
\hline
\end{longtable}

\begin{longtable}{rcccccccc}
\caption[]{The abundances of lithium, carbon, nitrogen, oxygen, sodium and
 magnesium in atmospheres of studied stars}\\
\hline
   HD & \Teff &\logg &(Li/H) &(C/H) & (N/H)& (O/H) &(Na/H) &(Mg/H) \\
\hline
\endfirsthead
\caption[]{Continued.}\\
\hline
   HD & \Teff &\logg &(Li/H) &(C/H) & (N/H)& (O/H) &(Na/H) &(Mg/H) \\
\hline
\endhead
\hline
\endfoot
\label{tableA4}
  2910 &   4756 &   2.70 &   --      &   8.42 &   8.25 &   8.80 &   6.30 &   7.45  \\
  4188 &   4809 &   2.70 &   --      &   8.32 &   8.20 &   8.85 &   6.28 &   7.41  \\
  4482 &   4917 &   2.65 &     0.65  &   8.25 &   8.20 &   8.60 &   6.32 &   7.55  \\
  5395 &   4849 &   2.15 &   --      &   8.05 &    --  &   8.50 &   5.80 &   7.43  \\
  6319 &   4650 &   2.30 &   --      &   8.30 &   8.10 &   8.72 &   6.31 &   7.55  \\
  6482 &   4738 &   2.40 &   --      &   8.20 &   8.05 &   8.63 &   6.18 &   7.47  \\
  7106 &   4684 &   2.55 &   --      &   8.40 &   8.20 &   8.80 &   6.32 &   7.56  \\
  7578 &   4680 &   2.50 &   --      &   8.40 &   8.40 &   8.75 &   6.70 &   7.58  \\
  8207 &   4750 &   2.75 &   --      &   8.35 &   8.40 &   8.80 &   6.47 &   7.60  \\
  8599 &   4781 &   2.50 &   --      &   8.15 &   8.00 &   8.63 &   6.04 &   7.47  \\
  8733 &   4932 &   2.70 &   1.05    &   8.25 &   8.10 &   8.57 &   6.20 &   7.50  \\
  9408 &   4804 &   2.30 &   --      &   8.22 &   7.95 &   8.70 &   6.00 &   7.35  \\
 10975 &   4881 &   2.20 &   --      &   8.15 &   7.95 &   8.65 &   6.08 &   7.40  \\
 11559 &   4977 &   3.00 &   --      &   8.30 &   8.30 &   8.80 &   6.36 &   7.54  \\
 11749 &   4679 &   2.40 &   --      &   8.25 &   7.80 &   8.80 &   6.05 &   7.42  \\
 11949 &   4708 &   2.30 &   --      &   8.13 &   7.95 &   8.60 &   6.12 &   7.48  \\
 15453 &   4696 &   2.40 &   0.90  &   8.25 &   8.00 &   8.72 &   6.23 &   7.52  \\
 15755 &   4611 &   2.30 &   --      &   8.25 &   8.15 &   8.75 &   6.22 &   7.56  \\
 15779 &   4821 &   2.70 &   --      &   8.30 &   8.25 &   8.78 &   6.28 &   7.43  \\
 16247 &   4629 &   2.20 &   --      &   8.20 &    --  &   8.63 &   5.98 &   7.43  \\
 16400 &   4840 &   2.50 &   --      &   8.26 &   8.25 &   8.72 &   6.32 &   7.40  \\
 17361 &   4646 &   2.50 &   --      &   8.37 &   8.15 &   8.75 &   6.37 &   7.55  \\
 18885 &   4722 &   2.50 &   --      &   8.40 &   8.28 &   8.80 &   6.51 &   7.55  \\
 19270 &   4723 &   2.40 &   --      &   8.30 &   8.30 &   8.68 &   6.49 &   7.58  \\
 19787 &   4832 &   2.75 &   --      &   8.35 &   8.23 &   8.75 &   6.41 &   7.53  \\
 19845 &   4933 &   2.80 &   --      &   8.30 &   8.30 &   8.65 &   6.47 &   7.50  \\
 20791 &   4986 &   2.80 &   --      &   8.33 &   8.25 &   8.75 &   6.40 &   7.48  \\
 25602 &   4693 &   2.40 &   --      &   8.08 &   7.80 &    --  &   6.02 &   7.40  \\
 25604 &   4764 &   2.70 &   --      &   8.40 &   8.35 &   8.80 &   6.47 &   7.52  \\
 26546 &   4743 &   2.25 &   --      &   8.22 &   8.10 &   8.65 &   6.40 &   7.50  \\
 26659 &   5178 &   2.90 &   --      &    --  &    --  &    --  &   6.26 &   7.33  \\
 26755 &   4630 &   2.20 &   --      &    --  &    --  &    --  &   6.42 &   7.55  \\
 27348 &   5003 &   2.80 &   --      &    --  &    --  &    --  &   6.43 &   7.53  \\
 27371 &   4955 &   2.70 &   0.90  &   8.35 &   8.40 &   8.72 &   6.66 &   7.60  \\
 27697 &   4975 &   2.65 &   0.80  &   8.25 &   8.50 &   8.60 &   6.60 &   7.58  \\
 28292 &   4453 &   2.10 &   --      &   8.30 &   8.00 &   8.70 &   6.18 &   7.45  \\
 28305 &   4925 &   2.55 &   0.70  &   8.35 &   8.40 &   8.80 &   6.67 &   7.60  \\
 28307 &   4961 &   2.70 &   1.10  &   8.35 &   8.35 &   8.70 &   6.54 &   7.50  \\
 30557 &   4829 &   2.45 &   --      &   8.15 &   8.10 &   8.65 &   6.26 &   7.35  \\
 31444 &   5080 &   2.75 &   0.60  &   8.25 &   8.10 &   8.60 &   6.22 &   7.37  \\
 33419  &  4708 &   2.30 &    --      &   8.30 &   8.20 &   8.75 &   6.48 &   7.55  \\
 33618  &  4590 &   2.30 &    --      &   8.42 &   8.23 &   8.78 &   6.54 &   7.62  \\
 34200  &  5055 &   2.80 &    --      &    --  &    --  &    --  &   6.28 &   7.49  \\
 34559  &  5010 &   2.90 &    --      &   8.35 &   8.30 &   8.80 &   6.40 &   7.45  \\
 35369  &  4931 &   2.40 &    --      &   8.10 &   8.10 &   8.67 &   6.10 &   7.35  \\
 37638  &  5093 &   2.80 &    --      &   8.25 &   8.25 &   8.72 &   6.25 &   7.48  \\
 39070  &  5047 &   2.80 &    --      &   8.25 &   8.30 &   8.72 &   6.28 &   7.49  \\
 39910  &  4618 &   2.60 &    --      &   8.60 &   8.25 &   8.90 &   6.61 &   7.61  \\
 40020  &  4670 &   2.30 &    --      &   8.30 &   8.40 &   8.72 &   6.62 &   7.59  \\
 40801  &  4703 &   2.20 &    --      &   8.18 &   7.90 &   8.55 &   6.05 &   7.50  \\
 42341  &  4655 &   2.60 &    0.85  &   8.44 &   8.40 &   8.80 &   6.75 &   7.65  \\
 43023  &  4994 &   2.40 &    --      &   8.15 &   8.12 &   8.57 &   6.22 &   7.37  \\
 45415  &  4762 &   2.30 &    --      &    --  &    --  &    --  &   6.17 &   7.43  \\
 46374  &  4661 &   2.30 &    0.85  &   8.25 &   8.20 &   8.70 &   6.23 &   7.43  \\
 46758  &  5003 &   2.90 &    --      &   8.22 &   8.10 &   8.72 &   6.22 &   7.25  \\
 47138  &  5211 &   3.00 &    1.10  &   8.20 &   8.25 &   8.72 &   6.26 &   7.40  \\
 47366  &  4772 &   2.60 &    --      &   8.20 &   7.95 &   8.65 &   6.19 &   7.37  \\
 48432  &  4836 &   2.65 &    --      &   8.25 &   8.15 &   8.80 &   6.12 &   7.37  \\
 50904  &  4953 &   2.70 &    --      &   8.20 &   8.20 &   8.60 &   6.21 &   7.37  \\
 53329  &  5012 &   2.80 &    --      &   8.25 &   7.95 &   8.85 &   5.95 &   7.28  \\
 54810  &  4669 &   2.40 &    --      &   8.20 &   7.85 &   8.65 &   5.98 &   7.30  \\
 55280  &  4654 &   2.25 &    --      &   8.25 &   8.16 &   8.65 &   6.35 &   7.48  \\
 56891  &  4709 &   2.40 &    --      &   8.30 &   8.25 &   8.70 &   6.45 &   7.55  \\
 58207  &  4799 &   2.35 &    --      &   8.22 &   8.05 &   8.65 &   6.21 &   7.46  \\
 59686  &  4654 &   2.40 &    --      &   8.37 &   8.25 &   8.85 &   6.44 &   7.55  \\
 60294  &  4569 &   2.15 &    --      &   8.22 &   8.15 &   8.60 &   6.45 &   7.47  \\
 60318  &  4962 &   2.80 &    --      &   8.32 &   8.30 &   8.77 &   6.50 &   7.45  \\
 60341  &  4634 &   2.15 &    --      &   8.35 &   8.20 &   8.80 &   6.47 &   7.55  \\
 60986  &  5057 &   2.60 &    --      &    --  &    --  &    --  &   6.50 &   7.50  \\
 61363  &  4785 &   2.10 &    --      &   8.10 &   8.05 &   8.60 &   6.10 &   7.40  \\
 61935  &  4780 &   2.40 &    --      &   8.25 &   8.20 &   8.72 &   6.30 &   7.43  \\
 62141  &  4971 &   2.80 &    0.95  &   8.25 &   8.05 &   8.60 &   6.20 &   7.42  \\
 62345  &  5032 &   2.60 &    --      &   8.30 &   8.30 &   8.60 &   6.48 &   7.48  \\
 63798  &  5004 &   2.50 &    1.75  &   8.28 &   8.15 &   8.60 &   6.28 &   7.42  \\
 64152  &  4977 &   2.70 &    --      &   8.30 &   8.20 &   8.60 &   6.44 &   7.50  \\
 64967  &  4864 &   2.55 &    --     &    --  &    --  &   8.45 &   5.75 &   7.12  \\
 65066  &  4868 &   2.60 &    --      &    --  &    --  &    --  &   6.34 &   7.46  \\
 65345  &  4963 &   2.70 &    --      &   8.23 &   8.10 &   8.72 &   6.41 &   7.41  \\
 67539  &  4781 &   2.45 &    --      &    --  &   7.75 &   8.60 &   5.75 &   7.27  \\
 68312  &  5090 &   2.70 &    --      &   8.25 &   8.30 &   8.72 &   6.26 &   7.42  \\
 68375  &  5071 &   2.90 &   --     &    8.35 &   8.25 &   8.90 &   6.26 &   7.45  \\
 70523  &  4642 &   2.20 &   --     &     --  &    --  &    --  &   6.04 &   7.48  \\
 71088  &  4900 &   2.70 &   --     &    8.30 &   8.05 &   8.76 &   6.17 &   7.35  \\
 73017  &  4693 &   2.30 &   --     &     --  &    --  &    --  &   5.76 &   7.15  \\
 74794  &  4701 &   2.25 &   --     &    8.37 &   8.15 &   8.72 &   6.37 &   7.57  \\
 75506  &  4876 &   2.50 &   --     &    8.20 &   8.10 &   8.85 &   6.02 &   7.35  \\
 75958  &  5030 &   2.70 &   --     &    8.25 &   8.15 &   8.72 &   6.14 &   7.35  \\
 76291  &  4495 &   2.00 &   --     &    8.22 &   7.90 &   8.60 &   6.18 &   7.50  \\
 76813  &  5060 &   2.80 &   --     &    8.22 &   8.25 &   8.75 &   6.28 &   7.45  \\
 78235  &  5070 &   2.80 &   --     &    8.20 &   8.25 &   8.67 &   6.36 &   7.40  \\
 79181  &  4867 &   2.40 &   --     &    8.20 &   8.00 &   8.70 &   6.10 &   7.38  \\
 80546  &  4601 &   2.25 &   --     &    8.28 &   8.15 &   8.62 &   6.35 &   7.50  \\
 81688  &  4789 &   2.30 &   --     &    8.20 &   7.85 &   8.65 &   5.94 &   7.33  \\
 82969  &  4948 &   2.70 &   --     &    8.18 &   8.05 &   8.60 &   6.00 &   7.40  \\
 83240  &  4682 &   2.45 &    1.10&    8.25 &   8.20 &   8.75 &   6.20 &   7.37  \\
 83371  &  4861 &   2.60 &   --     &    8.15 &   7.95 &   8.55 &   5.85 &   7.35  \\
 86513  &  4755 &   2.30 &   --     &    8.25 &   8.20 &   8.72 &   6.30 &   7.50  \\
 90633  &  4596 &   2.30 &    1.85&     --  &    --  &    --  &   6.37 &   7.60  \\
 93291  &  5061 &   2.75 &   --     &    8.28 &   8.15 &   8.65 &   6.30 &   7.41  \\
 93875  &  4590 &   2.25 &   --     &    8.43 &   8.25 &   8.76 &   6.47 &   7.58  \\
 94084  &  4787 &   2.65 &   --     &    8.42 &   8.35 &   8.87 &   6.55 &   7.50  \\
 94402  &  5004 &   2.70 &   --     &    8.30 &   8.30 &   8.72 &   6.43 &   7.52  \\
 94497  &  4702 &   2.30 &   --     &    8.20 &   7.95 &   8.65 &   6.13 &   7.38  \\
 95808  &  4946 &   2.55 &   --     &    8.20 &   8.30 &   8.67 &   6.44 &   7.47  \\
 98366  &  4702 &   2.40 &   --     &    8.22 &   8.00 &   8.57 &   6.18 &   7.43  \\
 100696 &  4862 &   2.40 &   --     &    8.22 &   8.00 &   8.77 &   5.98 &   7.30  \\
 101484 &  4991 &   2.70 &   --     &    8.25 &   8.15 &   8.72 &   6.36 &   7.50  \\
 102928 &  4654 &   2.35 &   --     &    8.25 &   8.05 &   8.73 &   6.13 &   7.40  \\
 103605 &  4611 &   2.35 &   --     &    8.42 &   8.05 &   8.85 &   6.18 &   7.59  \\
 103912 &  4870 &   2.80 &   --     &     --  &    --  &    --  &   5.52 &   7.27  \\
 104783 &  5247 &   2.55 &   --     &     --  &    --  &    --  &   5.80 &   7.23  \\
 106714 &  4935 &   2.50 &   --     &    8.20 &   8.15 &   8.80 &   6.19 &   7.39  \\
 108381 &  4680 &   2.50 &   --     &    8.42 &   8.60 &   8.72 &   6.71 &   7.65  \\
 109053 &  4921 &   2.50 &   --     &    8.00 &   7.80 &   8.50 &   5.93 &   7.38  \\
 110024 &  4921 &   2.70 &   --     &    8.25 &   8.20 &   8.68 &   6.36 &   7.50  \\
 113321 &  4739 &   2.10 &   --     &    8.30 &   8.10 &   8.72 &   6.23 &   7.62 \\
 113997 &  4697 &   1.70 &   --     &    8.30 &   8.10 &   8.60 &   6.41 &   7.35 \\
 114357 &  4551 &   2.30 &   --     &    8.40 &   8.40 &   8.80 &   6.63 &   7.60  \\
 116292 &  4922 &   2.60 &    1.40&     --  &    --  &    --  &   6.20 &   7.48  \\
 117304 &  4630 &   2.17 &    1.00&    8.15 &   8.00 &   8.60 &   6.10 &   7.40  \\
 117566 &  5475 &   3.15 &   --     &     --  &    --  &    --  &   6.38 &   7.48  \\
 119126 &  4802 &   2.25 &   --     &    8.15 &   8.15 &   8.70 &   6.25 &   7.40  \\
 120084 &  4883 &   2.55 &   --     &    8.33 &   8.15 &   8.65 &   6.37 &   7.47  \\
 120164 &  4746 &   2.30 &   --     &  8.20 &   8.15 &   8.72 &   6.25  &  7.40  \\
 120420 &  4676 &   2.15 &   --     &  8.10 &   7.90 &   8.60 &   5.98  &  7.37  \\
 136138 &  4995 &   2.60 &    1.30&  8.18 &   8.15 &   8.55 &   6.20  &  7.36  \\
 138852 &  4859 &   2.30 &   --     &  8.17 &   7.95 &   8.72 &   6.07  &  7.38  \\
 139254 &  4708 &   2.35 &    1.00&  8.25 &   8.10 &   8.70 &   6.23  &  7.45  \\
 139329 &  4690 &   2.30 &   --     &   --  &   8.05 &   8.75 &   6.13  &  7.45  \\
 143553 &  4644 &   2.30 &   --     &  8.10 &    --  &   8.40 &   5.94  &  7.27  \\
 146388 &  4731 &   2.45 &   --     &   --  &    --  &    --  &   6.52  &  7.59  \\
 148604 &  5110 &   2.80 &    0.90&   --  &    --  &    --  &   6.16  &  7.25  \\
 152224 &  4685 &   2.25 &   --     &  8.10 &   7.95 &   8.62 &   6.06  &  7.37  \\
 153956 &  4604 &   2.45 &   --     &  8.54 &    --  &   8.95 &   6.64  &  7.62  \\
 155970 &  4717 &   2.50 &   --     &  8.35 &   8.20 &   8.65 &   6.50  &  7.55  \\
 156874 &  4881 &   2.50 &   --     &  8.20 &   8.20 &   8.60 &   6.23  &  7.45  \\
 159353 &  4850 &   2.40 &   --     &  8.20 &   8.30 &   8.65 &   6.20  &  7.50  \\
 161178 &  4789 &   2.20 &   --     &  8.22 &   8.05 &   8.60 &   6.15  &  7.50  \\
 162076 &  4959 &   2.70 &    1.00&  8.25 &   8.25 &   8.60 &   6.32  &  7.42  \\
 166578 &  4859 &   2.50 &   --     &  8.10 &   7.95 &   8.55 &   5.71  &  7.25  \\
 168653 &  4632 &   2.20 &   --     &  8.18 &   7.95 &   8.60 &   6.10  &  7.40  \\
 170693 &  4256 &   1.50 &   --     &  8.00 &   7.70 &    --  &   5.76  &  7.23  \\
 171994 &  5014 &   2.70 &    0.60&  8.20 &   8.05 &   8.58 &   6.15  &  7.33  \\
 175743 &  4669 &   2.50 &   --     &   --  &    --  &    --  &   6.28  &  7.48  \\
 176408 &  4564 &   2.25 &   --     &  8.32 &   8.30 &   8.72 &   6.50  &  7.54  \\
 176598 &  5024 &   2.80 &    1.20&  8.25 &   8.38 &   8.60 &   6.45  &  7.47  \\
 180711 &  4824 &   2.40 &   --     &  8.15 &   8.10 &   8.67 &   6.16  &  7.35  \\
 185644 &  4591 &   2.40 &   --     &  8.28 &   8.10 &   8.70 &   6.33  &  7.60  \\
 187739 &  4649 &   2.30 &   --     &  8.00 &   --   &   8.45 &   5.98  &  7.13  \\
 188119 &  4993 &   2.75 &   --     &  8.20 &   8.30 &   8.85 &   6.00  &  7.28  \\
 192787 &  4987 &   2.60 &   --     &  8.20 &   8.20 &   8.65 &   6.21  &  7.40  \\
 192836 &  4772 &   2.60 &    1.25&  8.32 &   8.15 &   8.72 &   6.31  &  7.47  \\
 195330 &  4792 &   2.40 &   --     &   --  &    --  &    --  &   6.12  &  7.10  \\
 196134 &  4741 &   2.40 &   --     &  8.15 &   7.95 &   8.55 &   6.06  &  7.33  \\
 198431 &  4524 &   2.00 &   --     &   --  &   7.85 &   8.70 &   6.02  &  7.38  \\
 199870 &  4937 &   2.70 &   --     &  8.28 &   8.20 &   8.65 &   6.46  &  7.53  \\
 204771 &  4904 &   2.70 &   --     &  8.25 &   8.30 &   8.75 &   6.38  &  7.55  \\
 206005 &  4709 &   2.20 &   --     &  8.15 &   8.00 &   8.60 &   6.24  &  7.45  \\
 207130 &  4792 &   2.60 &   --    &   --  &    --  &    --  &   6.55  &  7.55  \\
 208111 &  4592 &   2.30 &   --     &  8.43 &   8.75 &   8.75 &   6.58  &  7.65  \\
 211006 &  4553 &   2.35 &   --     &  8.42 &   8.20 &   8.72 &   6.54  &  7.58  \\
 212496 &  4646 &   2.30 &   --     &  8.06 &   7.85 &   8.42 &   5.95  &  7.25  \\
 214567 &  4981 &   2.50 &   --     &  8.15 &   8.15 &   8.65 &   6.22  &  7.39  \\
 215030 &  4723 & 2.35 & --   &  8.00 &   --   &   8.37 &   5.81  &  7.15  \\
 215721 &  4890 & 2.40 & --   &  8.10 &   8.00 &   8.55 &   5.85  &  7.27  \\
 216131 &  4984 & 2.70 & --   &  8.22 &   8.20 &    --  &   6.27  &  7.42  \\
 216228 &  4698 & 2.40 & --   &  8.22 &   8.15 &   8.68 &   6.27  &  7.34  \\
 218031 &  4692 & 2.20 & --   &  8.20 &   8.00 &   8.70 &   6.04  &  7.45  \\
 219418 &  5281 & 2.80 & --   &  8.15 &   8.35 &   8.60 &   6.25  &  7.42  \\
 219916 &  5038 & 2.80 & --   &  8.00 &   8.15 &   8.65 &   6.31  &  7.45  \\
 221345 &  4664 & 2.20 & --   &   --  &    --  &    --  &   6.00  &  7.50  \\
 221833 &  4603 & 2.30 & --   &  8.37 &   8.00 &   8.72 &   6.38  &  7.50  \\
 225197 &  4734 & 2.50 & --   &  8.37 &   8.40 &   8.75 &   6.55  &  7.60  \\
 225216 &  4720 & 2.20 & --   &  8.20 &   8.20 &   8.67 &   6.29  &  7.50  \\
 BD+22 2606&4680& 2.10&  --  &  8.15&    8.00&    8.60&  6.28& 7.49 \\
 BD+25 2555&5014& 2.80&  --  &  8.10&    7.70&    9.00&  5.86& 7.44 \\
 BD+28 2250&4630& 2.00&  --  &  --  &    --  &    8.75&  5.76& 7.24 \\
\hline
\hline
\end{longtable}


\begin{thebibliography}{}
\bibitem[1994]{Alex}
 Alexander D. R., Ferguson J. W. 1994, ApJ, 437, 879
\bibitem[2005]{alfet05}
 Affer, L., Micela, G., Morel, T., Sanz-Forcada, J., Favata, F. 2005, A\&A
 433, 647
\bibitem[1999]{AAM99}
 Alonso A., Arribas S., Mart\'{i}nez-Roger C. 1999, A\&AS 139, 335
\bibitem[2002]{andet02}
 Andrievsky S.M., Egorova I.A., Korotin S.A., Burnage R., 2002, A\&A 389, 519
\bibitem[1999]{Ang99}
 Angulo C., Arnould M., Rayet M. et al., 1999,
  Nucl. Phys. A656, 3
\bibitem[2006]{benfel06}
 Bensby F., Feltzing S. 2006, MNRAS 367, 1181.
\bibitem[2006]{bienet06}
 Bienaym\'e, O., Soubiran, C., Mishenina, T.V.,  Kovtyukh V.V.,
  Sibert, A. 2006, A\&A 446, 933
\bibitem[1998]{BLG98}
 Blackwell, D.E. \& Lynas--Gray, A.E. 1998, A\&AS 129, 505
\bibitem[1996]{boyet96}
 Boyarchuk A.A., Antipova L.I., Boyarchuk M.E. et al. 1996, Astron.
  Zhurn. 73, 862
\bibitem[2001]{boyet01}
 Boyarchuk A.A., Antipova L.I., Boyarchuk M.E. et al. 2001, Astron.
  Zhurn. 78, 349
\bibitem[1989]{broet89}
 Brown I.A., Sneden C., Lambert D.L., Dutchover E. Jr., 1989 ApJS 71 293
\bibitem[1986]{car86}
 Carlsson M.  Uppsala Obs. Rep., 1986, 33
\bibitem[1970]{cayet70}
 Cayrel de Strobel G., Chave-Godard J., Hernandez G. et al. 1970,
   A\&A, 7, 408
\bibitem[1996]{cayet96}
 Cayrel R., Faurobert-Scholl M., Feautrier N., et al. 1996,
   A\&A, 312, 549.
\bibitem[1994]{cc94}
 Charbonnel, C., 1994, A\&A 282, 811
\bibitem[1998]{charet98}
 Charbonnel, C., Brawn J.A., Wallerstein G., 1998, A\&A 332, 204
\bibitem[2000]{charbal00}
 Charbonnel, C., Balachandran S. C., 2000, A\&A 359, 563
\bibitem[2005]{ccst05}
 Charbonnel, C., Talon, S., 2005, Science 309, 2189
\bibitem[2000]{delyi00}
 Deliyannis C.P., Pinsonneault M.H., Charbonnel C., 2000, IAU Symp. 198.
 Eds Da Sylva L., de Medeiros R. \& Spite M., p.61
\bibitem[1988]{edv88}
 Edvardsson B. , 1988, A\&A 190, 148
\bibitem[1993]{edv93}
 Edvardsson B., Andersen J., Gustafsson B. et al. 1993,
 A\&A 275, 101
 \bibitem[ESA, 1997]{esa97}
 ESA 1997, The Hipparcos and Tycho Catalogues, (Noordwijk)
 Series: ESA-SP 1200, Netherlands: ESA Publications Division
\bibitem[1992]{gal92}
 Galazutdinov G.A., 1992, Preprint SAO RAS, n92
\bibitem[2001]{gray01}
 Gray D.F., Brown K., 2001, PASP, 113, 723
\bibitem[1999]{gir99}
 Girardi L., 1999, MNRAS, 308, 818
\bibitem[2001]{hale01}
 Hale S.E., Champagne A.E., Illiadis C., Hansper V.Y., Powell D.C.,
 Blackmon J.C., 2001, Physical Review C, 65, 015801
\bibitem[2004]{hale04}
 Hale S.E., Champagne A.E., Illiadis C., Hansper V.Y., Powell D.C.,
 Blackmon J.C., 2004, Physical Review C, 70, 045802
\bibitem[1998]{har98}
 Harmanec, P., 1998,  A\&A, 335, 173
\bibitem[1969]{Hu69} Hubbard W. B., Lampe M. 1969, ApJS, 18, 297
\bibitem[1975]{Iben75} Iben I. Jr., 1975, ApJ 196, 525
\bibitem[1991]{iben91}
 Iben I., 1991, ApJ SS, 76, 55
\bibitem[1996]{IR96} Iglesias C. A., Rogers F. J. 1996, ApJ 464, 943
\bibitem[1983]{Ito83} Itoh N., Mitake S., Iyetomi H. Ichimaru S. 1983, ApJ,
 273, 774
\bibitem[2003]{katzet03}
 Katz, D., Farata, F., Aigrain, S., Micela, G. 2003, A\&A 397, 747
\bibitem[1998]{katzet98}
 Katz, D., Soubiran, C., Cayrel, R., Adda, M. \& Cautain, R.
  1998, A\&A, 338, 151
\bibitem[1982]{kijgus82}
 Kj$\ae$rgaard P., Gustafsson B. 1982, A\&A 115, 145
\bibitem[1999a]{koret99a}
 Korotin S.A., Andrievsky S.M., Kostynchuk L.Yu., 1999a,
  Astron. Sp. Sci.,  260, 531
\bibitem[1999b]{koret99b}
 Korotin, S. A.; Andrievsky, S. M.; Luck, R. E. 1999b,
  A\&A, 351,  168.
\bibitem[1999]{kormish99}
 Korotin S.A., Mishenina T.V., 1999, Astron. Zhurn., 76, 611
\bibitem[1999]{kovand99}
 Kovtyukh V.V., Andrievsky S.M. 1999, A\&A, 351, 597
\bibitem[2006]{kovtet06}
 Kovtyukh V.V.,  Mishenina T.V., Gorbaneva T.I., Bienaym\'e O.,
  Soubiran C.,  Kantcen L.E. 2006 Astron. Reports. 50, 134
\bibitem[1993]{kur93}
 Kurucz R.L., 1993, CD ROM n13
\bibitem[1996]{kuzsha96}
 Kuznetsova L.A., Shavrina A.V., 1996, KNFT, 12, 75
\bibitem[1981]{lamrie81}
 Lambert D. L., Ries L. M., 1981, ApJ,  248, 228
\bibitem[1999]{mal99}
 Mallik S.V., 1999, A\&A 352, 495
\bibitem[1993]{maset93}
 Mashonkina, L.I., Sakhibullin, N.A., Shimanskii, V.V. 1993,
 AZh 70, 372
\bibitem[1997]{meret97}
 Mermilliod J.C., Hauck B., Mermilliod M.,  1997, A\&AS 124, 349
\bibitem[1997]{mistsy97}
 Mishenina T.V., Tsymbal V.V., 1997 Pis'ma v Astron. Zhurn., 23, 693
\bibitem[2004]{mskk04}
 Mishenina, T.V., Soubiran, C., Kovtyukh, V.V., Korotin, S.A.
 2004, A\&A, 418, 551
\bibitem[2003]{mkks03}
 Mishenina, T. V.; Kovtyukh, V. V.; Korotin, S. A.; Soubiran, C.
 2003, AZh 80, 458
\bibitem[1984]{Mi84}
 Mitake S., Ichimaru S., Itoh N. 1984, ApJ 277, 375
\bibitem[2006]{palaciosetal06}
Palacios A., Charbonnel C., Talon S., Siess L., 2006, A\&A in press,
 astro-ph/0602389
\bibitem[2004]{paset04}
 Pasquini L., Randich, S., Zoccalli, M., Hill V., Charbonnel, C., Nordstrom, B.
 2004, A\&A 424, 951
\bibitem[2006]{reddet06}
 Reddy B.E., Lambert D.L., Prieto C.A. 2006, MNRAS 367, 1329
\bibitem[1975]{Reimers75} Reimers, D., 1975, Mem. Soc. Roy. Sci. Li\`ege,
 6th Ser., 8, 369
\bibitem[1987]{sah87}
 Sakhibullin, N.A. 1987, AZh 64, 1269
\bibitem[2003]{siebet03}
 Siebert, A., Bienaym\'e, O., Soubiran, C. 2003, A\&A 399, 531
\bibitem[2000]{siess00}
 Siess L., Dufour E., Forestini M., 2000, A\&A 358, 593
\bibitem[2003]{soubet03}
 Soubiran, C., Bienaym\'e, O., Siebert, A. 2003, A\&A 398, 141
\bibitem[2005]{sougir05}
 Soubiran C.,  Girard P.,  2005 A\&A 438, 139
\bibitem[1998]{soubet98}
 Soubiran, C., Katz, D. \& Cayrel, R., 1998, A\&AS, 133, 221
\bibitem[2000]{ss00}
 Strassmeier K.G., Schordan P., 2000, Astron. Nachricht.
  321, 277
\bibitem[1996]{tsy96}
 Tsymbal V.V. 1996 ASP Conf. Ser. 108, 198
\end{thebibliography}
\end{document}